\documentclass[useAMS,usenatbib,usegraphicx,onecolumn]{mn2e}

\newcommand{\vg}{\bmath{g}}
\newcommand{\vv}{\bmath{v}}
\newcommand{\vx}{\bmath{x}}
\newcommand{\vOmega}{\mathbf\Omega}
\newcommand{\btimes}{\bmath{\times}}
\newcommand{\bdot}{\bmath{\cdot}}
\newcommand{\dd}{\mathrm{d}}
\newcommand{\ee}{\mathrm{e}}

\begin{document}

\title{Continuous-wave gravitational radiation from pulsar glitch recovery}
\author[M.~F.~Bennett, C.~A.~van Eysden and A.~Melatos]{M.~F.~Bennett,\thanks{Email: mfb@unimelb.edu.au} C.~A.~van Eysden and A.~Melatos \\
School of Physics, University of Melbourne, Parkville, VIC 3010, Australia}

\date{Accepted XXXX. Received XXXX; in original form XXXX}
\pagerange{\pageref{firstpage}--\pageref{lastpage}} \pubyear{2010}

\maketitle

\label{firstpage}

\begin{abstract}
Nonaxisymmetric, meridional circulation inside a neutron star, excited by a glitch and persisting throughout the post-glitch relaxation phase, emits gravitational radiation.  Here, it is shown that the current quadrupole contributes more strongly to the gravitational wave signal than the mass quadrupole evaluated in previous work.  We calculate the signal-to-noise ratio for a coherent search and conclude that a large glitch may be detectable by second-generation interferometers like the Laser Interferometer Gravitational-Wave Observatory.  It is shown that the viscosity and compressibility of bulk nuclear matter, as well as the stratification length-scale and inclination angle of the star, can be inferred from a gravitational wave detection in principle.
\end{abstract}

\section{Introduction}
\label{sec:introduction}
Rotation-powered radio pulsars are promising sources of high frequency gravitational waves.  Their spin frequencies often lie in the hectohertz `sweet spot' of current detectors, e.g. the Laser Interferometer Gravitational-Wave Observatory (LIGO).  The rotation of their crusts can be  measured extremely precisely, enabling coherent searches which improve the signal-to-noise ratio by the square root of the number of wave cycles observed.  Such coherent searches have already beaten electromagnetic spin-down limits on the quadrupole moment of the Crab \citep{abb08} and are close for other pulsars \citep{abb07}.  There are two main obstacles to detection.  (1) Dephasing occurs if the radio pulses are used to construct a gravitational wave phase model but the fluid interior rotates at a slightly different speed to the crust.  (2) The quadrupoles predicted so far are relatively small in isolated pulsars without any ongoing accretion activity, e.g. unstable oscillations such as r-modes \citep{bri04, nay06, bon07}, precession \citep{jon02b}, internal magnetic deformations \citep{bon96, cut02}, quasiradial fluctuations \citep{sed03, sid09}, and hydrodynamic turbulence \citep{mel10}.  Accreting millisecond pulsars can reach larger quadrupoles through magnetically confined mountains \citep{mel05, pay06, vig09} or thermal mountains \citep{ush00, has06}.

In this paper, we investigate another source of gravitational radiation from isolated pulsars, namely the radiation emitted during the recovery phase following a pulsar glitch \citep{van08}.  Glitches are small, abrupt jumps $\Delta \nu$ in the rotation frequency $\nu$ which range in fractional size from $10^{-11}$ to $10^{-4}$ across the pulsar population and over four decades in individual objects.  Currently, out of $\sim 1800$ known pulsars, 101 have been observed to glitch, with a total of 285 individual events \citep{mel08}.  Glitches occur randomly in all but two objects (PSR J0537$-$6910 and PSR J0835$-$4510), which spin up quasiperiodically \citep{mel08}.  Most pulsars which have glitched at all have only glitched once.  Of the 35 per cent that have glitched multiple times, and with the exception of the quasiperiodic pair, the glitch sizes and waiting times are well fitted by power-law and Poissonian probability density functions respectively \citep{mel08}, consistent with an avalanche mechanism \citep{war08, mel09}.

Most theories of pulsar glitches build on the vortex unpinning paradigm introduced by \citet{and75}.  Superfluid vortices pin to lattice sites or defects in the crust and are prevented from migrating outward as the crust spins down electromagnetically.  At some stage, many vortices unpin catastrophically, transferring angular momentum to the crust.  While it is unknown what triggers the collective unpinning, it is likely to excite a nonaxisymmetric flow for two generic reasons.  (1) Pinning causes the crust and superfluid to rotate differentially, inevitably driving nonaxisymmetric meridional circulation and even turbulence, as observed in laboratory experiments \citep{mun75, nak83, jun00} and numerical simulations \citep{per05, per06a, per06b, mel07, per08, per09} of spherical Couette flow.  (2) Avalanche trigger mechanisms, like self-organized criticality, which are favoured by the observed glitch statistics, intrinsically lead to an inhomogeneous and hence nonaxisymmetric superfluid velocity field, with spatial fluctuations correlated on all scales, from the smallest to the largest \citep{jen98, mel08}.

The gravitational wave signal from a pulsar glitch separates into two parts.  First, there is a burst corresponding to nonaxisymmetric vortex unpinning and rearrangement during the spin-up event itself.  To date, observations have failed to resolve the spin-up time-scale.  In the Vela pulsar, which was monitored continuously for several years, it occurs over less than 40~s \citep{mcc90, dod02}.  Second, there is a decaying continuous-wave signal during the quasi-exponential relaxation phase (lasting days to weeks) following the spin-up event \citep{she96}.  The latter signal arises as viscous interactions between the crustal lattice and core superfluid erase the nonaxisymmetry in the superfluid velocity field and restore the crust and core to co-rotation (or at least steady differential rotation).  \citet{sid09} constructed a two-fluid `body-averaged' model of a glitch and calculated that the burst signal emitted during the spin-up event by coupling to quasiradial oscillations is too weak to be detected.  In this paper, we focus on the second part of the signal, which has the advantage of enduring for many rotation periods, enabling a coherent search with increased signal to noise.

Two techniques have been proposed to date to search for gravitational radiation emitted during the spin-up event and post-glitch relaxation.  \citet{cla07} developed a Bayesian selection criterion for comparing f-mode ringdown to white noise.  \citet{hay08} investigated coherent network analysis, which does not assume any particular waveform.  Both methods would be aided by the availability of a specific signal template, like the one calculated in this paper.  Importantly, by combining such a template with data, gravitational wave experiments can constrain the equation of state of bulk nuclear matter, complementing particle accelerator experiments which have recently produced results that disagree with astrophysical data.  Heavy ion and nuclear resonance experiments measuring the compressibility of nuclear matter imply a soft equation of state \citep{stu01,vre03}, whereas neutron star observations imply a hard equation of state, albeit at lower energies \citep{har06, lat07}.  Likewise, heavy-ion colliders measure a viscosity close to the conjectured quantum lower bound \citep{ada07}, whereas the relaxation time-scale of pulsar glitches suggests a value many orders of magnitude larger \citep{cut87,and05,van10}.  Gravitational wave observations will help to resolve these and other issues; bulk matter at nuclear density cannot be assembled in terrestrial laboratories with current technology \citep{van08, owe09, xu09}.  

In this paper, we calculate the gravitational radiation generated from the spin up of the stellar interior following a pulsar glitch.  We estimate its detectability with the current generation of long-baseline interferometers, and show that certain important constitutive properties of a neutron star can be extracted from gravitational wave data, at least in principle.  The calculation is based on \citet{van08}, extended to treat current quadrupole radiation.  In Section~\ref{sec:ekmanflow},  we solve the general hydrodynamic problem of nonaxisymmetric, stratified, compressible spin-up flow in a cylinder, driven by Ekman pumping, following an abrupt increase in the angular velocity of the container.  The initial and boundary conditions implemented by \citet{van08} are modified slightly to make them more realistic.  In Section~\ref{sec:gravitationalwavecalculation} we predict the gravitational radiation emitted during the relaxation phase following a glitch.  We calculate the signal-to-noise ratio and estimate the detectability of the signal in Section~\ref{sec:detectability}.  In Section~\ref{sec:nuclearproperties},  we show how to extract the compressibility, stratification, and viscosity of the stellar interior from gravitational wave data.

\section{Ekman flow following a glitch} \label{sec:ekmanflow}

Radio pulse timing experiments have so far failed to resolve temporally the abrupt increase in the angular velocity of the neutron star crust during a glitch \citep{mcc90, dod02}.  Hence, in the absence of more detailed information, we model a glitch as a step increase in the angular velocity $\Omega$ of a rotating, rigid, cylindrical container filled with a Newtonian fluid \citep{abn96, van08}.   A cylinder is a coarse approximation to a spherical star, but it admits analytic solutions and has a long history of being used to model neutron stars and in geomechanical studies \citep{ped67, wal69, abn96, van08}.

Differential rotation between the container and interior fluid drives Ekman pumping, which spins up the interior over time; see \citet{ben74} for a review of Ekman pumping.  The spin up of an axisymmetric container was first treated analytically by \citet{gre63}.  For an incompressible fluid, the entire volume is spun up on the Ekman time-scale, $t_E=E^{-1/2}\Omega^{-1}$, where $E=\nu/(\Omega L^2)$ defines the dimensionless Ekman number in terms of the kinematic viscosity $\nu$ and the size $L$ of the container.  Subsequently, it was shown that compressibility and stratification reduce the spun-up volume by hindering flow along the side walls \citep{wal69, abn96, van08}.  With less volume to spin up, the Ekman time-scale is lower.  Nonaxisymmetric spin up was analysed by \citet{van08}.

In this section, we solve the problem of the nonaxisymmetric, stratified, compressible spin up of a cylinder, extending  \citet{van08}.  We write down the linearised hydrodynamic equations  in Section~\ref{subsec:modelequations}, solve for the general spin-up flow in Section~\ref{subsec:spinupflow}, apply initial and boundary conditions in Section~\ref{subsec:boundaryconditions} and \ref{subsec:initialconditions}, and discuss precisely how and why these conditions differ from previous analyses.  The final, time-dependent solutions for the pressure, density and velocity fields are presented in Section~\ref{subsec:solutions}.  We discuss the initial conditions for a glitch in Section~\ref{subsec:chooseBCICforglitch}.  For full details of the calculation, the reader is referred to Section~2 of \citet{van08}.

\subsection{Model equations} \label{subsec:modelequations}

Consider a cylinder of height $2L$ and radius $L$, containing a compressible, Newtonian fluid with uniform kinematic viscosity $\nu$, and rotating about the $z$ axis with angular velocity $\vOmega=\Omega\bmath{e_z}$.  In the rotating frame, the compressible Navier-Stokes equation reads
{\setlength{\arraycolsep}{0pt}
\begin{equation}
	\frac{\partial \vv}{\partial t} + \vv \bdot \nabla \vv + 2\vOmega\btimes\vv =  - \frac{1}{\rho} \nabla p + \vg+ \nu\nabla^2 \vv+\frac{\nu}{3}\nabla\left( \nabla\bdot\vv\right) + \nabla\left(\frac{1}{2}\Omega^2r^2\right)~.
\label{eq:navier-stokes}
\end{equation}
}
The fluid satisfies the continuity equation
\begin{equation}
	\frac{\partial \rho}{\partial t} + \nabla \bdot (\rho \vv) = 0~,
\end{equation}
and the energy equation is written in a form that relates the convective derivatives of the pressure and density,
\begin{equation}
 \left( \frac{\partial }{\partial t} + \vv\bdot\nabla \right)\rho  = \frac{1}{c^{2}}\left( \frac{\partial }{\partial t} + \vv\bdot\nabla  \right) p~.
 \label{eq:energyequation}
\end{equation}
The symbols $\vv$, $\rho$, $p$, $\vg$ and $c$ represent the fluid velocity, density, pressure, gravitational acceleration, and the speed of sound, which is determined by the equation of state.  Following \citet{abn96}, gravity is taken to be uniform and directed towards the midplane of the cylinder, 
\begin{equation}
	\vg =  \left\{ \begin{array}{ll} -g \bmath{e_z} & \textrm{if $z>0$}~,\\ +g \bmath{e_z} & \textrm{if $z<0$}~.
\end{array} \right.
\end{equation}
where $g$ is constant.

We work in cylindrical coordinates $(r,\phi,z)$ and consider the region $z \ge 0$, as the flow is symmetric about $z=0$.  Equations (\ref{eq:navier-stokes})--(\ref{eq:energyequation}) are rewritten in dimensionless form by making the substitutions $t \mapsto t_E t$, $r \mapsto L r$, $z \mapsto L z$, $\vv \mapsto L \delta\Omega \vv$, $\rho \mapsto \rho_0 \rho$, $p \mapsto \rho_0 g L p$, and $\nabla \mapsto ({1}/{L}) \nabla$, where the scale factor $\rho_0$ is chosen to be the equilibrium density at $z=0$.  The scaled equations obtained in this way [see equations (6)--(8) in \citet{van08}] feature three dimensionless quantities: the Rossby number $\epsilon=\delta\Omega / \Omega$, the Froude number $F=L\Omega^2/g$, and the scaled compressibility $K=gL/c^2$.

\subsection{Spin-up flow} \label{subsec:spinupflow}

At time $t=0$, the angular velocity of the cylinder accelerates instantaneously from $\Omega$ to $\Omega+\delta\Omega$.  If $\epsilon$ is small, as in a pulsar glitch, the problem linearises and we can solve for the equilibrium and spin-up flows separately by making the  
perturbation expansions $\rho \mapsto \rho + \epsilon \delta\rho$, $p \mapsto p + \epsilon \delta p$, and $\vv \mapsto \bmath{\delta v}$.  In the frame rotating at $\Omega$, the equilibrium velocity is zero and the spin-up flow is of order $\epsilon$.

We assume the equilibrium state is steady and axisymmetric, with $\rho=\rho(r,z)$ and $p=p(r,z)$.  Ignoring centrifugal terms proportional to $F$, and taking $\rho^{-1} d\rho/d z $ to be uniform for simplicity, as in previous work \citep{wal69, abn96, van08}, we find
\begin{eqnarray}
	\rho(z)&=&\ee^{-K_s z}~, \label{eq:rhoequilsolution}\\
	p(z)&=&K_s^{-1} \ee^{-K_s z}~ \label{eq:pequilsolution},
\end{eqnarray}
where $K_s=L/z_s=-L \rho^{-1} d\rho/d z $ is a constant which depends on the stratification length-scale, $z_s$.

The spin-up flow is unsteady and nonaxisymmetric, with $\delta\rho = \delta\rho(r,\phi,z,t)$, $\delta p = \delta p(r,\phi,z,t)$, and $\bmath{\delta v}=\bmath{\delta v}(r,\phi,z,t)$.  We solve equations (17)--(21) in \citet{van08} for the spin-up flow using the method of multiple scales, expanding $\bmath{\delta v}$, $\delta p$ and $\delta \rho $ as perturbation series in the small parameter $E^{1/2}$, e.g. $\delta\rho=\delta\rho^0+E^{1/2}\delta\rho^1+O(E)$ \citep{wal69, abn96, van08}.  Following Section~2.3 in \citet{van08}, the $O(E^0)$ continuity equation is automatically satisfied and the order $O(E^{1/2})$ equations reduce to
\begin{equation}
	\frac{1}{r} \frac{\partial}{\partial r} \left( r \frac{\partial \Phi}{\partial r} \right) + \frac{1}{r^2} \frac{\partial^2 \Phi}{\partial \phi^2} - \frac{4 K_s}{N^2} \frac{\partial \Phi}{\partial z} + \frac{4}{N^2} \frac{\partial^2 \Phi}{\partial z^2}=0~,
 \label{eq:masterphiequation}
\end{equation}
where $N^2=\left(K_s-K\right)/F$ is the dimensionless Brunt-V\"ais\"al\"a frequency and we define $\Phi = - \partial \left( \delta p^0/\rho\right)/ \partial t$.

Equation (\ref{eq:masterphiequation}) can be solved by separation of variables.  The general solution that is regular as $r\to 0$ has the form
\begin{equation}
	\Phi(r,\phi,z,t) = F \sum^{\infty}_{m=0} J_m(\lambda r) \left[ A_{m}\cos(m\phi)+B_{m}\sin(m\phi)\right] Z_{m}(z) T_{m}(t)~,
 \label{eq:generalsolution1}
\end{equation}
where $m\geq 0$ is an integer and $\lambda$ is determined by the boundary conditions.  The prefactor $F$ is included as $\Phi$ is expected to be of this order.  This is the same result found by \citet{van08} but is slightly more general than the equivalent in \citet{abn96}, as it allows for the possibility that $F N^2$ and $K$ are of similar magnitude, a likely scenario in a neutron star \citep{van08}.

\subsection{Boundary conditions
 \label{subsec:boundaryconditions}}

The boundary conditions on $\Phi$ are set by the boundary conditions on the $O(E^0)$ velocity fields,
\begin{eqnarray}
	\delta v_r^0 &=& - \frac{1}{2Fr} \frac{\partial}{\partial \phi} \left( \frac{\delta p^0}{\rho} \right)~,\label{eq:orderE0vr}\\
	\delta v_\phi^0 &=& \frac{1}{2F} \frac{\partial}{\partial r} \left( \frac{\delta p^0}{\rho} \right)~,\label{eq:orderE0vphi}
\end{eqnarray}
as $\Phi$ is defined in terms of $\delta p^0$ and is therefore $O(E^0)$ too.  [To impose boundary conditions on the $O(E^{1/2})$ flow, we would need to know $\delta p^1$.]  Assuming no penetration at the side wall, we have $\partial \Phi /\partial \phi =0$ at $r=1$ and hence
\begin{equation}
	\Phi(r,\phi,z,t) = F \sum_{m=0}^{\infty} \sum_{n=1}^\infty J_m(\lambda_{mn} r) \left[ A_{mn}\cos(m\phi)+B_{mn}\sin(m\phi)\right] Z_{mn}(z) T_{mn}(t)~,
 \label{eq:generalsolution2}
\end{equation}
where $\lambda_{mn}$ is the $n$th root of $J_m(\lambda)=0$.

To find $Z_{mn}$, we use the $O(E^{1/2})$ axial flow,
\begin{equation}
	\delta v_z^1 = \frac{1}{FN^2} \frac{\partial \Phi}{\partial z} - \Phi~,\label{eq:orderE1vz}
\end{equation}
as $\delta v_z^0 = 0$.  We require $\delta v_z^1=0$ at $z=0$, so that the flow is symmetric about the midplane.  The normalisation of $Z_{mn}$ is arbitrary, and we choose $Z_{mn}(1)=1$, giving
\begin{equation}
 Z_{mn}(z)=\frac{\left(F N^2 - \beta_{-} \right) \ee^{\beta_{+}z} - \left(F N^2-\beta_{+}\right) \ee^{\beta_{-}z}}{\left(F N^2-\beta_{-}\right)\ee^{\beta_{+}}-\left(F N^2-\beta_{+}\right)\ee^{\beta_{-}}}~,
 \label{eq:Zmn}
\end{equation}
with
\begin{equation}
 \beta_\pm =\frac{1}{2}\left[ K_{s} \pm \left( K_{s}^2+N^2 \lambda^2_{mn}\right)^{1/2}\right]~.
 \label{eq:betaplusminus}
\end{equation}

Another boundary condition applies to the top and bottom faces of the cylinder, which determines $T_{mn}$.  The mass flux into and out of the Ekman boundary layer at $z=\pm1$ is related to the circulation just outside this layer by \citep{ped67, wal69, abn96, van08}
\begin{equation}
	\left. \delta v_z \right|_{z=\pm 1} = \left. \mp \frac{1}{2} E^{1/2} \left[ \nabla \btimes \right( \bmath{\delta v} - \bmath{v_B} \left) \right]_z \right|_{z=\pm 1}~,
 \label{eq:deltavzmassflux}
\end{equation}
where $\bmath{v_B}$ is the dimensionless velocity of the boundary in the frame rotating at $\Omega$.  Ekman pumping continues until the local fluid velocity, here $\bmath{\delta v}$, matches the boundary velocity $\bmath{v_B}$.  For a rigid container, the final angular velocity equals $\Omega+\delta\Omega$ in the inertial observer's frame, corresponding to $\bmath{v_B} = r \bmath{e_\phi}$ in the rotating frame.  To find $T_{mn}$, we differentiate (\ref{eq:deltavzmassflux}) with respect to time and substitute equation (\ref{eq:orderE1vz}) into the left hand side of (\ref{eq:deltavzmassflux}) (note: $\delta v_z^0 =0$), and equations (\ref{eq:orderE0vr}) and (\ref{eq:orderE0vphi}) into the right hand side of (\ref{eq:deltavzmassflux}).  After some algebra, we find that the $(m,n)$-th mode relaxes exponentially as $\Phi \propto \exp(-\omega_{mn} t)$, with
\begin{equation}
	\omega_{mn} = \frac{\lambda_{mn}^2\left[\left(F N^2 -\beta_{-}\right)\ee^{\beta_{+}}-\left(F N^2-\beta_{+}\right)\ee^{\beta_{-}}\right]}{\left(4 F K+\lambda_{mn}^2\right)\left(\ee^{\beta_{+}}-\ee^{\beta_{-}}\right)}~.
\label{eq:omega_mn}
\end{equation}

Integrating $\Phi$ with respect to time, the general solution for the pressure perturbation can be written as
\begin{equation}
	\frac{\delta p^0 (r,\phi,z,t)}{\rho(z)} = C(r,\phi,z) + F \sum_{m=0}^\infty \sum_{n=1}^\infty J_m(\lambda_{mn} r) \left( A_{mn}\cos{m\phi}+B_{mn}\sin{m\phi}\right) Z_{mn}(z) \ee^{-\omega_{mn} t}~,
 \label{eq:pressuresolution1}
\end{equation}
where $A_{mn}$ and $B_{mn}$ absorb a factor of $\omega_{mn}^{-1}$, and $C(r,\phi,z)$ is the constant of integration.  $C(r,\phi,z)$ is constrained by the boundary condition (\ref{eq:deltavzmassflux}) and must match the boundary velocity at $z=1$.  Using (\ref{eq:orderE0vr}) and (\ref{eq:orderE0vphi}), we obtain
\begin{eqnarray}
	{v_B}_r(r,\phi,1) &=& -\frac{1}{2 F r} \frac{\partial C(r,\phi, 1)}{\partial \phi} \label{eq:Ccondition1}~,\\
	{v_B}_\phi(r,\phi,1) &=& \frac{1}{2 F} \frac{\partial C(r,\phi, 1)}{\partial r} \label{eq:Ccondition2}~.
\end{eqnarray}

\subsection{Initial conditions
 \label{subsec:initialconditions}}

All that remains is to specify the initial conditions, which determine $A_{mn}$ and $B_{mn}$.  Without specialising to a particular trigger for the spin-up event at $t=0$ or modelling the vortex unpinning and rearrangement that presumably accompanies it, we consider the general situation where these processes establish some instantaneously nonaxisymmetric pressure field throughout the interior.  [Five possible physical causes of the nonaxisymmetry are discussed in detail in Section~1 of \citet{van08}.]  We denote the initial state at $t=0$ by the symbol $\delta P^0(r,\phi,z) = \delta p (r,\phi,z,0) / \rho(z)$.  Specifying $\delta P^0(r,\phi,z)$ is equivalent to specifying the initial velocity or density, which are related through (\ref{eq:orderE0vr}), (\ref{eq:orderE0vphi}), (\ref{eq:orderE1vz}), and the $O(E^0)$ equation of motion,
\begin{equation}
	\delta\rho^0 = -\frac{\partial \delta p^0}{\partial z}~.\label{eq:orderE0rho}
\end{equation}
The choice of $\delta P^0(r,\phi,z)$ is arbitrary, but it should satisfy the boundary conditions outlined in Section~\ref{subsec:boundaryconditions}.  We eliminate $C(r,\phi,z)$ by evaluating (\ref{eq:pressuresolution1}) at $t=0$, obtaining
\begin{equation}
	\frac{\delta p^0 (r,\phi,z,t)}{\rho(z)} = \delta P^0(r,\phi,z) + F \sum_{m=0}^\infty \sum_{n=1}^\infty J_m(\lambda_{mn} r) \left(A_{mn}\cos{m\phi}+B_{mn}\sin{m\phi}\right) Z_{mn}(z) \left( \ee^{-\omega_{mn} t} - 1 \right)~.
	\label{eq:pressuresolution2}
\end{equation}
The coefficients $A_{mn}$ and $B_{mn}$ are determined at $z=1$ from $\delta P^0(r,\phi,z)$ and $C(r,\phi,1)$ .  In general, we have
\begin{equation}
	A_{mn} = \frac{2}{\pi F J_{m+1}^2(\lambda_{mn})} \int_0^{2\pi} d\phi \int_0^1 \dd r \, r J_m(\lambda_{mn} r) \cos(m\phi) \left[ \delta P^0(r,\phi,1) - C(r,\phi,1) \right]~.\label{eq:Amn}
\end{equation}
$B_{mn}$ is given by the same formula, with $\cos(m\phi)$ replaced by $\sin(m\phi)$.

\subsection{Velocity, density, and pressure solutions \label{subsec:solutions}}
Equations (\ref{eq:orderE0vr}), (\ref{eq:orderE0vphi}), (\ref{eq:orderE1vz}), (\ref{eq:orderE0rho}), and (\ref{eq:pressuresolution2}) yield complete solutions for the velocity, density and pressure fields.  Upon transforming back to dimensional variables and out of the rotating frame into the inertial observer's frame, we can write the results as follows:
\begin{eqnarray}
	v_r(r,\phi,z,t) &=& \delta v_r(r,\phi,z,0) + \frac{1}{2} L^2 \delta\Omega \sum_{m=0}^\infty \sum_{n=1}^\infty \frac{m}{r} J_m(\lambda_{mn}r/L) \left[ \frac{\left(F N^2-\beta_- \right) \ee^{\beta_+ z/L}-\left( F N^2-\beta_+\right) \ee^{\beta_- z/L}}{\left(F N^2-\beta_-\right) \ee^{\beta_+}-\left(F N^2-\beta_+\right)\ee^{\beta_-}} \right]\nonumber \\
	&&{}\qquad\qquad\qquad\qquad\qquad\qquad\qquad \times \left\{ A_{mn}\sin[m(\phi-\Omega t)] - B_{mn} \cos[m(\phi-\Omega t)] \right\} \left( \ee^{-E^{1/2}\omega_{mn} \Omega t} - 1 \right) ~,
 \label{eq:vrsolution}
\end{eqnarray}
\begin{eqnarray}
	v_\phi(r,\phi,z,t) &=& \Omega r + \delta v_\phi(r,\phi,z,0) + \frac{1}{2} L \delta\Omega \sum_{m=0}^\infty \sum_{n=1}^\infty \lambda_{mn} J_m'(\lambda_{mn}r/L) \left[ \frac{\left(F N^2-\beta_-\right) \ee^{\beta_+ z/L}-\left(F N^2-\beta_+\right) \ee^{\beta_- z/L}}{\left(F N^2-\beta_-\right) \ee^{\beta_+}-\left(F N^2-\beta_+\right)\ee^{\beta_-}} \right] \nonumber \\
	&&\qquad\qquad\qquad\qquad\qquad\qquad\qquad \times \left\{ A_{mn} \cos[m(\phi-\Omega t)] + B_{mn} \sin[m(\phi-\Omega t)] \right\} \left( \ee^{-E^{1/2}\omega_{mn} \Omega t} - 1 \right) ~,
 \label{eq:vphisolution}
\end{eqnarray}
\begin{eqnarray}
	v_z(r,\phi,z,t) &=& \frac{1}{4} L \delta\Omega E^{1/2} \sum_{m=0}^\infty \sum_{n=1}^\infty \lambda_{mn}^2
 J_m(\lambda_{mn}{r}/{L}) \left( \frac{\ee^{\beta_+ z/L}-\ee^{\beta_- z/L}}{\ee^{\beta_+}-\ee^{\beta_-}} \right) \nonumber \\
	&&\qquad\qquad\qquad\qquad\qquad\qquad\qquad \times \left\{ A_{mn} \cos[m(\phi-\Omega t)] + B_{mn} \sin[m(\phi-\Omega t)] \right\}  \left( \ee^{-E^{1/2}\omega_{mn} \Omega t} - 1 \right)~,
 \label{eq:vzsolution}
\end{eqnarray}
\begin{eqnarray}
	\rho(r,\phi,z,t) &=& \rho_0 \ee^{-z/z_s} + \delta\rho(r,\phi,z,0) + \frac{\rho_0 L \Omega \delta\Omega}{g} \sum_{m=0}^\infty \sum_{n=1}^\infty J_m(\lambda_{mn}{r}/{L}) \left[ \frac{\left(F N^2-\beta_- \right)\beta_- \ee^{-\beta_- z/L}-\left(F N^2-\beta_+ \right)\beta_+ \ee^{-\beta_+ z/L}}{\left(F N^2-\beta_-\right) \ee^{\beta_+}-\left(F N^2-\beta_+\right)\ee^{\beta_-}} \right] \nonumber \\
	&&\qquad\qquad\qquad\qquad\qquad\qquad\qquad \times \left\{ A_{mn} \cos[m(\phi-\Omega t)] + B_{mn} \sin[m(\phi-\Omega t)] \right\}  \left( \ee^{-E^{1/2}\omega_{mn} \Omega t} - 1 \right)~,
 \label{eq:rhosolution}
\end{eqnarray}
\begin{eqnarray}
	p(r,\phi,z,t) &=& \rho_0 g z_s e^{-z/z_s} + \delta p(r,\phi,z,0) + \rho_0 L^2 \Omega \delta\Omega \sum_{m=0}^\infty \sum_{n=1}^\infty J_m(\lambda_{mn}r/L) \left[ \frac{\left(F N^2-\beta_- \right)\ee^{-\beta_- z/L}-\left(F N^2-\beta_+ \right) \ee^{-\beta_+ z/L}}{\left(F N^2-\beta_- \right)\ee^{\beta_+}-\left(F N^2-\beta_+\right) \ee^{\beta_-}} \right] \nonumber \\
	&&\qquad\qquad\qquad\qquad\qquad\qquad\qquad \times \left\{ A_{mn} \cos[m(\phi-\Omega t)] + B_{mn} \sin[m(\phi-\Omega t)] \right\} \left( \ee^{-E^{1/2}\omega_{mn} \Omega t} - 1 \right) ~.
 \label{eq:psolution}
\end{eqnarray}
The initial velocity, density, and pressure are related to the chosen initial state $\delta P^0(r,\phi,z)$, through,
\begin{eqnarray}
	\delta v_r(r,\phi,z,0) &=& -\frac{1}{2Fr} \frac{\partial \delta P^0(r,\phi,z)}{\partial \phi}\\
	\delta v_\phi(r,\phi,z,0) &=& \frac{1}{2F} \frac{\partial \delta P^0(r,\phi,z)}{\partial r}\\
	\delta \rho(r,\phi,z,0) &=& - \frac{\partial \left[\rho(z) \delta P^0(r,\phi,z)\right]}{\partial z}\\
	\delta p(r,\phi,z,0) &=& \rho(z) \delta P^0(r,\phi,z)
\end{eqnarray}

In the limit $t\to\infty$, an incompressible fluid spins up completely via Ekman pumping and approaches a steady-state solution, which matches the boundary at $z=1$.  In contrast, for a compressible, stratified fluid, part of the volume is untouched by Ekman pumping.  In the latter case, the persistent, unaccelerated initial flow and the associated gradient in $v_\phi$ dissipate by viscous diffusion and adjust via inertial oscillations over the long time-scale $E^{-1} \Omega^{-1}$.

\subsection{$\delta P^0(r, \phi, z)$ for a glitch\label{subsec:chooseBCICforglitch}}
In this paper, we assume that a glitch spins up the crust rigidly and axisymmetrically but that it initially excites nonaxisymmetric motions in the fluid interior; that is, $\delta v_r^0$ and $\delta v_\phi^0$ are superpositions of $\cos(m\phi)$ and $\sin(m\phi)$ modes immediately after the glitch.  Possible physical mechanisms are outlined in Section~1 of \citet{van08}.  The crust spins up rigidly to angular velocity $\Omega + \delta\Omega$, which corresponds to $C(r,\phi,1)=Fr^2$ in equation (\ref{eq:Amn}), satisfying (\ref{eq:Ccondition1}) and (\ref{eq:Ccondition2}) as required.  The arbitrary initial pressure perturbation $\delta P^0(r,\phi,z)$, which specifies the initial flow velocity through (\ref{eq:orderE0vr}), (\ref{eq:orderE0vphi}) and (\ref{eq:orderE1vz}), is a sum of nonaxisymmetric modes satisfying the boundary conditions (e.g., no penetration of the side walls).  In dimensionless form, in the rotating frame, we can write
\begin{equation}
	\delta P^0(r,\phi,z) = F \sum_{m=1}^\infty C_m r^m \left( r^2 - 1 \right) \cos(m\phi)~.\label{eq:initialpressure}
\end{equation}
No $\sin(m\phi)$ terms or $z$ dependence are included for simplicity, and the relative weights of the modes are parametrized by the constants $C_m$.  We take $C_m=1$ for all $m$ in this paper.

The above initial condition is slightly more realistic than the one adopted by \citet{van08}, who posited that the perturbed (spin-up) flow develops from $\delta v_r^0 = \delta v_\phi^0 = 0$ immediately after the glitch to a permanently nonaxisymmetric steady-state flow at the boundary [see equations (40) and (41) in \citet{van08}].  There are two problems with the latter scenario.  First, it involves nonaxisymmetric, and therefore nonrigid, motion of the top and bottom faces of the cylindrical container, which in reality would exert large stresses on the stellar crust, probably causing it to crack.  Second, it artificially emits gravitational radiation in the steady state, even at $t \gg E^{-1} \Omega^{-1}$ (cf. Section~\ref{subsec:gravitationalwavestrain} below).

\section{Gravitational wave signal} \label{sec:gravitationalwavecalculation}
The gravitational radiation generated by the nonaxisymmetric spin-up flow in Section~\ref{sec:ekmanflow} is the sum of a mass quadrupole contribution, calculated previously by \citet{van08}, and a current quadrupole contribution.  The current quadrupole is typically smaller than the mass quadrupole by a factor $\sim c/v$.  However, using the results of Section \ref{sec:ekmanflow}, the nonaxisymmetric velocity perturbation is larger than the density perturbation by a factor $F$, implying a wave-strain ratio $h_\mathrm{mass}/h_\mathrm{current} \sim Fc/v \sim \Omega c/g$.  We compute the current quadrupole wave strain in this paper and refer to \citet{van08} for the mass quadrupole.

\subsection{Current quadrupole}

The far-field metric perturbation generated by a superposition of current multipole moments can be written as \citep{tho80}
\begin{equation}
	h_{jk}^{TT} = \frac{G}{c^5 D} \sum_{l=2}^\infty \sum_{m=-l}^l \frac{\partial^l S^{lm}(t)}{\partial t^l} T^{B2,lm}_{jk}~,
	\label{eq:gwstraindefinition}
\end{equation}
in the transverse, traceless gauge, where $t$ is the retarded time, $D$ is the distance from source to observer, and $T^{B2,lm}_{jk}$ is a tensor spherical harmonic which is a function of source orientation.  The $(l,m)$-th multipole moment, $S^{lm}(t)$, is given by \citep{mel10}
\begin{equation}
	S^{lm} = -\frac{32\pi}{(2l+1)!} \left[ \frac{l+2}{2l(l-1)(l+1)} \right]^{1/2} \int \dd^3 \vx\, r^l \vx \bdot \textrm{curl}(\rho \vv) Y^{lm*}~,
	\label{eq:currentquadrupoledefinition}
\end{equation}
for a Newtonian source, where $Y^{lm}$ denotes the usual scalar spherical harmonic.  In this paper, we only consider the leading order, quadrupole ($l=2$) term.  Importantly, $S^{2m}$ depends only on the Fourier mode with frequency $m\Omega$ in the spin-up flow described by equations (\ref{eq:vrsolution})--(\ref{eq:psolution}).  In other words, the $l=2$ metric perturbation is a linear superposition of terms each generated by a unique mode in the spin-up flow.

The plus and cross polarisations of the gravitational wave strain can be expressed compactly in terms of $S^{21}$ and $S^{22}$.  The axisymmetric Ekman flow leads to a quadrupole moment $\partial^2 S^{20}/\partial t^2 = O(E)$, which we neglect in this paper.  Denoting the inclination angle between the rotation axis of the star and the observer's line of sight by $i$, we can write
\begin{eqnarray}
	h_+(t) &=& \frac{G}{2 c^5 D} \left( \frac{5}{2\pi} \right)^{1/2} \left\{ \mathrm{Im}[\ddot{S}^{21}(t)] \sin i+ \mathrm{Im}[\ddot{S}^{22}(t)] \cos i \right\}~,\label{eq:hplus}\\
	h_\times(t) &=& \frac{G}{4 c^5 D} \left( \frac{5}{2\pi} \right)^{1/2} \left\{ \mathrm{Re}[\ddot{S}^{21}(t)] \sin 2i+ \mathrm{Re}[\ddot{S}^{22}(t)] (1+\cos^2 i) \right\}~,\label{eq:hcross}
\end{eqnarray}
where an overdot symbolises differentiation with respect to time.

\subsection{Gravitational wave strain\label{subsec:gravitationalwavestrain}}
We compute the far-field metric perturbation at a hypothetical detector by combining the flow solutions in Section~\ref{subsec:solutions} with the boundary and initial conditions in Section~\ref{subsec:chooseBCICforglitch}.  Appendix \ref{sec:appendixcurlrhov} shows how to rewrite the integral in (\ref{eq:currentquadrupoledefinition}) to involve only the pressure perturbation $\delta p^0$, simplifying the evaluation of $S^{lm}$.  The final result for the current quadrupole moment, for $0 < m \le 2$, takes the form
\begin{equation}
	S^{2m}(t) = \frac{(-1)^{m+1}8\pi(10\pi)^{1/2}}{15m}\rho_0 L^6 \delta\Omega \sum_{n=1}^\infty \left[ \left( U_{mn} - V_{mn} \right) \ee^{-i m \Omega t} + V_{mn} \ee^{-(E^{1/2}\omega_{mn} + i m) \Omega t} \right]~,\label{eq:S2m}
\end{equation}
with
\begin{eqnarray}
	U_{mn} &=& \delta_{n,1} \int_0^1 \dd r \int_0^1 \dd z \, r^{m+1} z^{2-m} \hat{U} r^m \left( r^2 - 1 \right) \ee^{-K_s z}~,\label{eqUmn}\\
	V_{mn} &=& \int_0^1 \dd r \int_0^1 \dd z \, r^{m+1} z^{2-m} \hat{U} A_{mn} J_m(\lambda_{mn} r) \left[ \frac{\left(F N^2-\beta_- \right)\ee^{-\beta_- z}-\left(F N^2-\beta_+ \right) \ee^{-\beta_+ z}}{\left(F N^2-\beta_- \right)\ee^{\beta_+}-\left(F N^2-\beta_+\right) \ee^{\beta_-}} \right],\label{eqVmn}
\end{eqnarray}
where
\begin{equation}
	\hat{U} = z \frac{\partial^2}{\partial r^2} + \frac{z}{r}\frac{\partial}{\partial r} - \frac{z m^2}{r^2} - r\frac{\partial^2}{\partial r \partial z} + 2 F \left( r^2 \frac{\partial^2}{\partial z^2} - rz\frac{\partial^2}{\partial r \partial z} - 2 z \frac{\partial}{\partial z} \right)
\end{equation}
is a differential operator acting on everything to its right in equations (\ref{eqUmn}) and (\ref{eqVmn}).  $U_{mn}$ and $V_{mn}$ are straightforward to calculate analytically, but the full expressions are too lengthy to quote here.

Substituting (\ref{eq:S2m}) into (\ref{eq:hplus}) and (\ref{eq:hcross}), we obtain the following expressions for the plus and cross polarisations as functions of time:
\begin{eqnarray}
	h_+(t) &=& h_0 \sum_{n=1}^\infty \Bigg[ \sin i \bigg\{ (U_{1n} - V_{1n})\sin\Omega t + V_{1n} \ee^{-E^{1/2}\omega_{1n} \Omega t} \left[ 2 E^{1/2}\omega_{1n}\cos\Omega t - \left( E \omega_{1n}^2 -1 \right) \sin\Omega t \right] \bigg\}\nonumber\\
	&&\qquad\qquad-\frac{1}{2} \cos i \bigg\{ 4 (U_{2n} - V_{2n})\sin 2\Omega t + V_{2n} \ee^{-E^{1/2}\omega_{2n}\Omega t} \left[ 4 E^{1/2}\omega_{2n}\cos 2\Omega t - \left( E \omega_{2n}^2 -4 \right) \sin 2\Omega t \right] \bigg\} \Bigg] ~, \label{eq:h_plus_final}\\
	h_\times(t) &=& \frac{h_0}{2} \sum_{n=1}^\infty \Bigg[ \sin 2i \Big\{ (V_{1n} - U_{1n})\cos\Omega t + V_{1n} \ee^{-E^{1/2}\omega_{1n}\Omega t} \left[ \left( E \omega_{1n}^2 -1 \right) \cos\Omega t + 2 E^{1/2}\omega_{1n}\sin\Omega t \right] \Big\} \nonumber\\
	&&\qquad\qquad - \frac{1}{2} (1+\cos^2 i) \bigg\{ 4(V_{2n} - U_{2n})\cos 2\Omega t + V_{2n} \ee^{-E^{1/2}\omega_{2n}\Omega t} \left[ \left( E \omega_{2n}^2 -4 \right) \cos 2\Omega t + 4 E^{1/2}\omega_{2n}\sin 2\Omega t \right] \bigg\} \Bigg]~,\label{eq:h_cross_final}
\end{eqnarray}
with
\begin{equation}
	h_0 = \frac{4\pi G \rho_0 L^6 \delta\Omega \, \Omega^2 }{3 c^5 D}\label{eq:h0}~.
\end{equation}
Equations (\ref{eq:h_plus_final}) and (\ref{eq:h_cross_final}) contain terms of order $(E \omega_{mn}^2)^0$, $(E \omega_{mn}^2)^{1/2}$ and $(E \omega_{mn}^2)^1$.  The derivation of the spin-up flow in Section~\ref{sec:ekmanflow} assumes $E^{1/2} \ll 1$.  Over the range of values for $K$, $N$ and $E$ that we consider in Section~\ref{sec:detectability} and \ref{sec:nuclearproperties}, it is also true that $E^{1/2} \omega_{m1} \ll 1$.  The quantity $E^{1/2} \omega_{mn}$ does become large for large $n$ ($\omega_{mn} \to n \pi / 2$ as $n\to\infty$), but the exponential suppresses the large-$n$ terms and the infinite sum converges.  For our purposes, truncating (\ref{eq:h_plus_final}) and (\ref{eq:h_cross_final}) at leading order $O(E^0)$ gives a good approximation.

In the scenario described in Section~\ref{subsec:chooseBCICforglitch}, the nonaxisymmetric initial perturbation is erased by Ekman pumping on the time-scale $t_E$, and the fluid spins up to rotate axisymmetrically with the boundary at $z=1$.  The effects of stratification and compressibility reduce the effectiveness of Ekman pumping, reducing the spin-up volume.  As a result, some regions of the interior are incompletely spun up and preserve some of their initial nonaxisymmetric flow for $t \gg t_E$, unlike in the incompressible problem.  The nonaxisymmetry persists, emitting gravitational radiation continuously, until viscous diffusion wipes it out on the time-scale $E^{-1} \Omega^{-1}$ \citep{gre63, ben74}. As the time-scale $E^{-1} \Omega^{-1} \ga 10^3$ years is comparable to, or greater than, the age of many glitching pulsars, one encounters the interesting possibility that neutron stars harbour a `fossil' nonaxisymmetric flow in their interior, preserved by stratification, which continually emits gravitational radiation, and whose structure reflects the history of differential rotation and superfluid vortex rearrangement in the star.  This possibility merits careful investigation in the future.  It is not the same as the artificial, nonaxisymmetric, nonrigid rotation of the crust postulated (for mathematical convenience) by \citet{van08} (cf. also Section~\ref{subsec:chooseBCICforglitch}).

\section{Detectability}\label{sec:detectability}

We now estimate the detectability of the gravitational wave signal derived in Section~\ref{sec:gravitationalwavecalculation} by calculating the signal-to-noise ratio expected to be achieved by current- and next-generation long-baseline interferometers.  The signal differs from a traditional continuous-wave source (e.g. an elliptical neutron star), because it decays over days to weeks (approximately $10^5$ to $10^8$ wave cycles).  It is therefore counterproductive to integrate coherently past a certain time (if one ignores the fossil quadrupole discussed in Section~\ref{subsec:gravitationalwavestrain}).  We find that the signal-to-noise ratio depends sensitively on the buoyancy, compressibility, and viscosity of the neutron star interior.  For certain, plausible ranges of these variables, the signal is detectable in principle by Advanced LIGO.

\subsection{Signal-to-noise ratio}\label{subsec:STNratio}
The response of a laser interferometer to plus and cross polarisations $h_+(t)$ and $h_\times(t)$ can be written as
\begin{equation}
	h(t) = F_+(t) h_+(t) + F_\times(t) h_\times(t)~.
\end{equation}
The beam-pattern functions $F_+$ and $F_\times$ depend on the rotation of the Earth and the sky position of the source \citep{jar98}.  For the signal in Section~\ref{sec:gravitationalwavecalculation}, it is convenient to split $h(t)$ into components that oscillate at the spin frequency of the star and its first harmonic, denoted $h_1(t)$ and $h_2(t)$ respectively.  Writing $h_{1,2}(t) = F_+(t) h_{{1,2}+}(t) + F_\times(t) h_{{1,2}\times}(t)$ and keeping terms of order $O(E^0)$ in (\ref{eq:h_plus_final}) and (\ref{eq:h_cross_final}), we find
\begin{eqnarray}
	h_{1+}(t) &=& h_0 \sin i \sin\Omega t \sum_{n=1}^\infty \left( U_{1n} - V_{1n} + V_{1n} \ee^{-E^{1/2}\omega_{1n} \Omega t} \right)~,\label{eq:h1plus}\\
	h_{2+}(t) &=& - 2 h_0 \cos i \sin 2\Omega t \sum_{n=1}^\infty \left( U_{2n} - V_{2n} + V_{2n} \ee^{-E^{1/2}\omega_{2n}\Omega t} \right)~,\label{eq:h2plus}\\
	h_{1\times}(t) &=& - \frac{h_0}{2} \sin 2i \cos\Omega t \sum_{n=1}^\infty \left( U_{1n} - V_{1n} + V_{1n}  \ee^{-E^{1/2}\omega_{1n}\Omega t} \right)~,\label{eq:h1cross}\\
	h_{2\times}(t) &=& h_0 (1+\cos^2 i) \cos 2\Omega t \sum_{n=1}^\infty \left( U_{2n} - V_{2n} + V_{2n} \ee^{-E^{1/2}\omega_{2n}\Omega t} \right)~.\label{eq:h2cross}
\end{eqnarray}

The signal-to-noise ratio $d$ for a quasi-dichromatic source (i.e. a source consisting of two narrow-band peaks at frequencies $f_*$ and $2f_*$) is given by equations (80)--(82) in \citet{jar98}.  The result is
\begin{equation}
	d^2 = \frac{2}{S_h(f_*)} \int_{-T_0/2}^{T_0/2} \dd t \left[ h_1(t) \right]^2 + \frac{2}{S_h(2f_*)} \int_{-T_0/2}^{T_0/2} \dd t \left[ h_2(t) \right]^2 ~.\label{eq:dsquared}
\end{equation}
In (\ref{eq:dsquared}), $S_h(f)$ is the spectral noise density of the interferometer at frequency $f$, $T_0$ denotes the total length of the coherent integration, and $f_* = \Omega / 2\pi$ is the stellar spin frequency.

The integration time for a coherent search is normally limited by computational expense rather than the length of the data stream.  Even when the radio ephemeris is known through radio observations, the radio and gravitational wave phases may not be equal, increasing the number of templates required for a search [e.g., the $\mathcal{F}$-statistic search for the Crab \citep{abb08}].  We assume a computational limit of two weeks for the remainder of this paper.  For the glitch recovery signal, the integration time is the minimum of the computational limit and the glitch recovery time-scale; integrating beyond the point where the signal decays away merely adds noise.  The exact value of $T_0$ which maximizes $d$ depends on the search algorithm, but it is always of order the $\ee^{-1}$ time constant for $h(t)$, i.e. $h(T_0)/h(0)=\ee^{-1}$.  For the general estimates below, we take $T_0 = (E^{1/2} \omega_{21} \Omega)^{-1}$, the $\ee^{-1}$ decay time-scale of the leading ($n=1$) term in equations (\ref{eq:h1plus})--(\ref{eq:h2cross}).  The $m=2$ mode decays more quickly than the $m=1$ mode, but the difference is moderate ($1 \le \omega_{21}/\omega_{11} \le 2$) over the parameter space that we consider.

\begin{figure*}
	\includegraphics{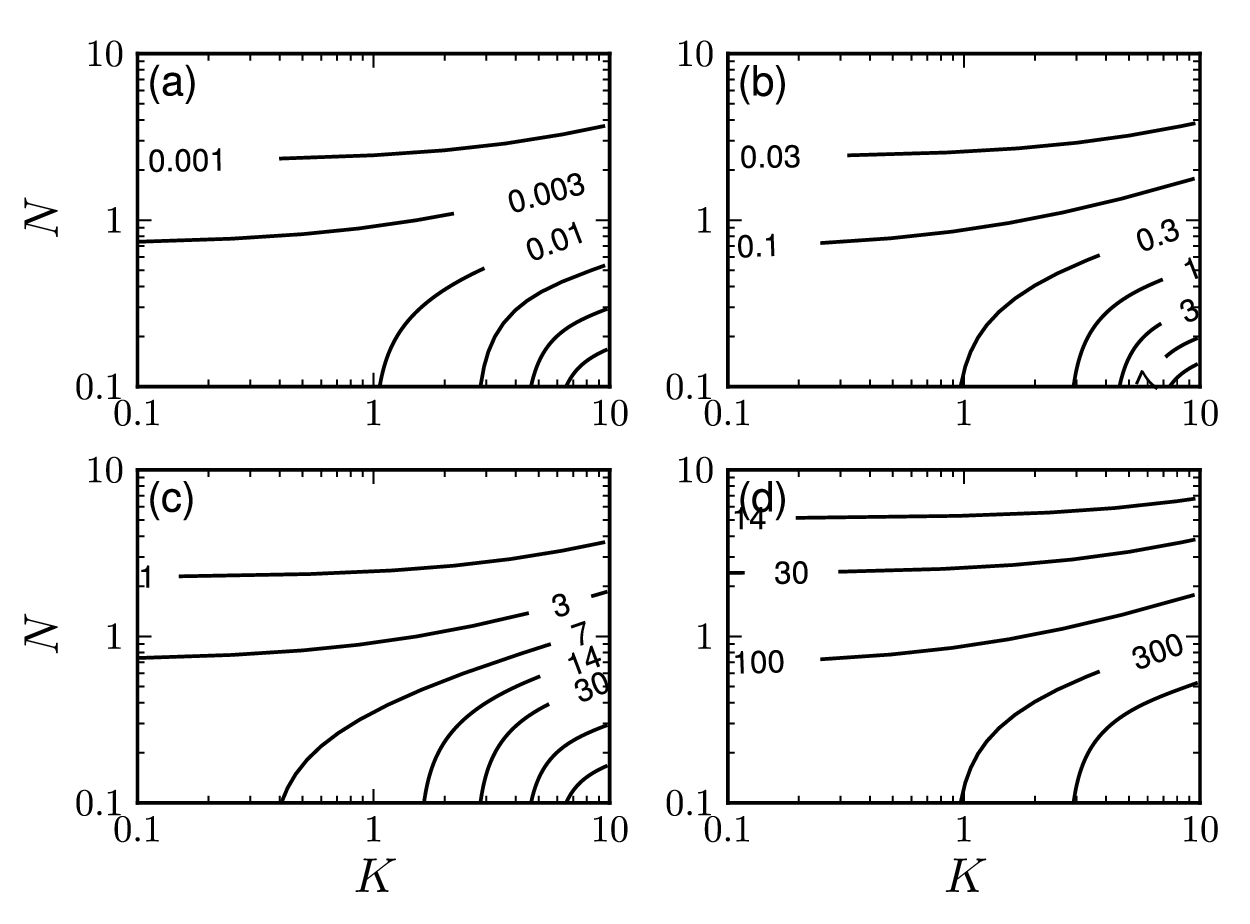}
	\caption{Contours of integration time $T_0$ (in days) as a function of the normalised compressibility $K$ and Brunt-V\"ais\"al\"a frequency $N$.  The integration time is chosen such that $h(t=T_0) = \mathrm{e}^{-1} h(t=0)$.  The Ekman number increases from the top left to the bottom right panels:  (a) $E = 10^{-11}$, (b) $E = 10^{-14}$, (c) $E = 10^{-17}$, (d) $E = 10^{-20}$.} \label{fig:integration_time}
\end{figure*}

Figure \ref{fig:integration_time} illustrates how $T_0$ depends on stellar parameters.  The four panels in Figure \ref{fig:integration_time} display contours of $T_0$ (in days) on the $K$-$N$ plane for four different values of $E$.  The value of $E$ in a neutron star is uncertain but Figure \ref{fig:integration_time} demonstrates that it plays a significant role in determining $T_0$.  One requires $E \sim 10^{-17}$ for the best match between $(E^{1/2}\omega_{21}\Omega)^{-1}$ and observed post-glitch recovery time-scales.  This value is artificially lower than that expected from neutron-neutron scattering, $E \sim 10^{-7}(\Omega/\textrm{rad s}^{-1})^{-1}$ \citep{cut87,and05,van10} because it is the effective value that arises when modelling the two-component Hall-Vinen-Bekarevich-Khalatnikov superfluid \citep{per05,and06} as a single Newtonian fluid \citep{eas79, abn96, van08}.

To calculate $d$, we evaluate (\ref{eq:dsquared}) with $T_0 = (E^{1/2}\omega_{21}\Omega)^{-1}$ and make several simplifying assumptions.  First, we approximate $h(t)$ by the leading ($n=1$) terms in the infinite sums in (\ref{eq:h1plus})--(\ref{eq:h2cross}).  For typical values of $N$ and $K$, this introduces an error of $\la 10$ per cent.  Second, following \citet{jar98}, we average the functions $\sin(m\Omega t)$ and $\cos(m\Omega t)$, which oscillate much more rapidly than $F_+$, $F_\times$, and $\exp(-t/T_0)$, over the observation period.  The result is
\begin{eqnarray}
	d^2 &=& \frac{(1-\ee^{-2}) h_0^2 A_1(K,N)}{S_h(f_*)} \int_0^{T_0}\dd t \left[ \sin^2 i \; F_+^2 + \frac{1}{4} \sin^2 2i \; F_\times^2 \right]\nonumber\\
	&& \qquad\qquad\qquad\qquad\qquad\qquad\qquad\qquad\qquad + \frac{(1-\ee^{-2}) h_0^2 A_2(K,N)}{S_h(2 f_*)}  \int_0^{T_0} \dd t \left[ 4\cos^2 i \; F_+^2 + (1+\cos^2 i)^2 \; F_\times^2 \right] ~, \label{eq:dsquareresult}
\end{eqnarray} 
with
\begin{equation}
	A_i(K,N) = \frac{1}{1-\ee^{-2}} (U_{i1}-V_{i1})^2 + \frac{2\ee}{1+\ee}(U_{i1}-V_{i1}) V_{i1} + \frac{1}{2} V_{i1}^2~.\label{eq:detectabilityAi}
\end{equation}
As discussed in Section~\ref{subsec:gravitationalwavestrain}, the signal is the sum of a persistent periodic signal associated with the fossil nonaxisymmetry (which decays on the long time-scale $E^{-1}\Omega^{-1} \gg T_0$) and the decaying signal generated by the Ekman flow.  To be conservative, we only consider the latter signal, setting $U_{mn}=V_{mn}$.  Hence, (\ref{eq:detectabilityAi}) reduces to $A_i(K,N)=V_{i1}^2/2$.

The signal-to-noise ratio depends on the right ascension $\alpha$, declination $\delta$, and polarisation angle $\psi$ of the source as well as the location and orientation of the interferometer and the diurnal phase of the Earth.  These quantities are usually known for any specific source.  However, to estimate detectability in general, we average $d$ over $\alpha$, $\delta$, $\psi$, and $i$ \citep{jar98}:
\begin{equation}
	\langle ... \rangle_{\alpha, \delta, \psi, i} = \frac{1}{2\pi} \int_0^{2\pi} \dd \alpha \times \frac{1}{2} \int_{-1}^1 \dd (\sin\delta) \times \frac{1}{2\pi} \int_0^{2\pi} \dd \psi \times \frac{1}{2} \int_{-1}^1 \dd (\cos i) \;(...)~.
\end{equation}
The beam pattern functions average to
\begin{equation}
	\left\langle \int_0^{T_0} \dd t F_+^2 \right\rangle_{\alpha, \delta, \psi} = \left\langle \int_0^{T_0} \dd t F_\times^2 \right\rangle_{\alpha, \delta, \psi} = \frac{T_0}{5} \sin^2\zeta~,\label{eq:averagebeampatternfunction}
\end{equation}
where $\zeta$ is the angle between the arms of the detector.  We give more details of this result in Appendix \ref{sec:appendixbeampatternfunctions}.  Substituting (\ref{eq:averagebeampatternfunction}) into (\ref{eq:dsquared}) and averaging over $i$, we obtain the following expression for the average signal-to-noise ratio:
\begin{equation}
	\left\langle d \right\rangle_{\alpha, \delta, \psi, i} = \frac{2}{5} \left(1-\ee^{-2}\right)^{1/2} h_0 \; T_0^{1/2} \sin\zeta \left[ \frac{A_1(K,N)}{S_h(f_*)} + \frac{4 A_2(K,N)}{S_h(2 f_*)} \right]^{1/2}~. \label{eq:signal-to-noise}
\end{equation}

\subsection{Second- and third-generation interferometers}\label{subsec:detectabilityestimate}

We now evaluate the signal-to-noise ratio (\ref{eq:signal-to-noise}) achieved by the second-generation interferometer LIGO, in both its Initial and Advanced configurations, and the third-generation, subterranean Einstein Telescope (ET).

\begin{figure*}
	\includegraphics{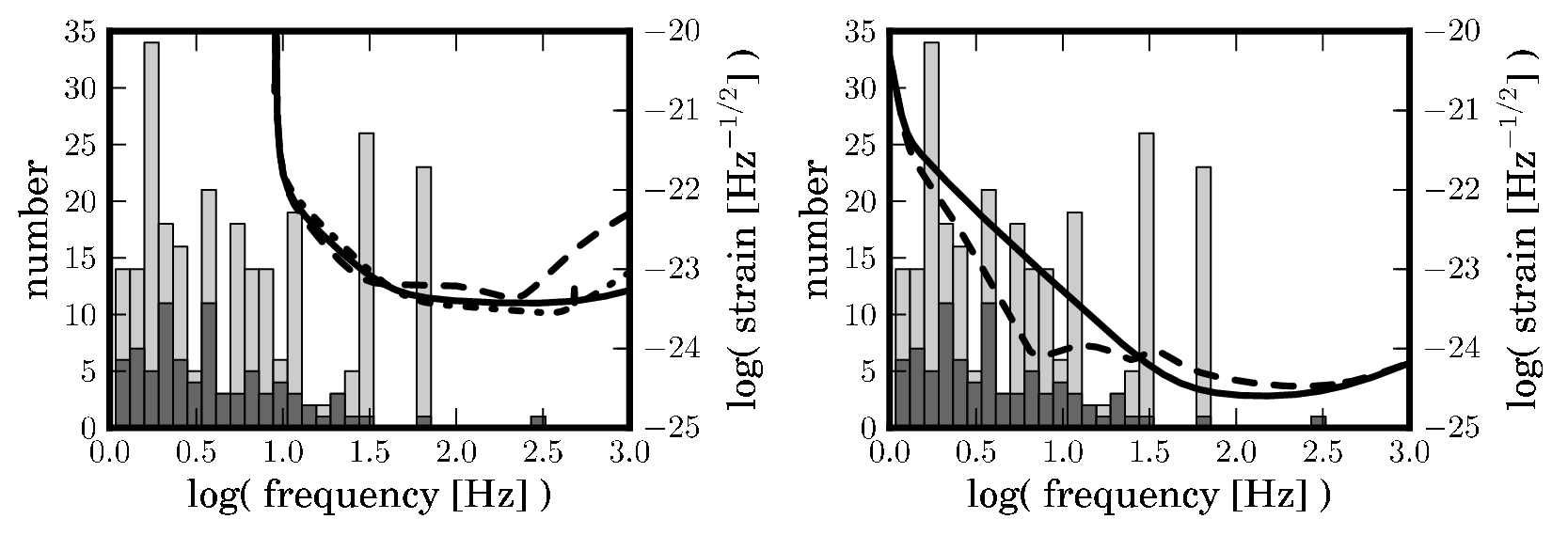}
	\caption{Histogram of known glitching pulsars (dark shading) and observed glitches (light shading) as a function of frequency.  Curves of anticipated spectral noise density for second- and third-generation interferometers are overlaid.  [Note: glitches emit gravitational radiation at both $f_*$ and $2f_*$ (see Section \ref{sec:gravitationalwavecalculation}).]  Left panel displays Advanced LIGO configurations: zero detuning, high power (solid), black hole optimised (dashed), neutron star optimised (dash-dotted).  Right panel displays ET configurations:  conventional (solid), xylophone (dashed).} \label{fig:hist}
\end{figure*}

There are various detector configurations proposed for Advanced LIGO\footnote{LIGO Document Control Center: document number LIGO-T0900288-v3}.  The best overall sensitivity across the entire frequency spectrum is achieved with zero detuning of the signal recycling mirror and high laser power.  Below 40 Hz, the configuration optimised for 30M$_\odot$ black hole binary inspirals provides the best sensitivity.  Above 40 Hz, the configuration optimised for 1.4M$_\odot$ neutron star binary inspirals provides the best sensitivity.  However, the differences between the three configurations are small.

Two configurations have been proposed for ET: a conventional interferometer \citep{hil08}, and a dual-band ‘xylophone’ configuration consisting of two co-located interferometers, one optimised for low frequencies and the other for high frequencies \citep{hil10}.  Below 30 Hz, the xylophone configuration is more sensitive than the conventional configuration, by a factor of up to $\sim 10$ in the 5–-10 Hz band.

Figure \ref{fig:hist} compares the spectral noise density of the different detector configurations.  It also bins
the number of known glitching pulsars and observed glitches as a function of frequency to illustrate which configurations are best suited for glitch searches.  It is important to recall that the results of Section \ref{sec:gravitationalwavecalculation} predict gravitational radiation at both the pulsar frequency $f_*$ and $2f_*$, with the pulsar orientation determining which frequency has the stronger signal.  The xylophone configuration of ET is the best choice for a glitch search.  More glitches have been observed in objects with $f_∗ < 30$ Hz (83 per cent), where the xylophone is more sensitive, than $f_∗ > 30$ Hz (17 per cent) \citep{per06,mel08}. Additionally, the increase in sensitivity of the ET xylophone configuration over the conventional configuration is far greater below 30 Hz than the decrease above 30 Hz.  In contrast, Advanced LIGO is not sensitive below 10 Hz and there is only a small difference in sensitivity between the different configurations over the frequency range where most glitches lie.  As mentioned above, the black-hole-optimised Advanced LIGO configuration is the most sensitive below 40 Hz but its advantage is slight and possibly outweighed by its slightly poorer performance at higher frequencies, where the strongest signals (from the fastest-spinning objects) arguably lie.

\begin{figure*}
	\includegraphics{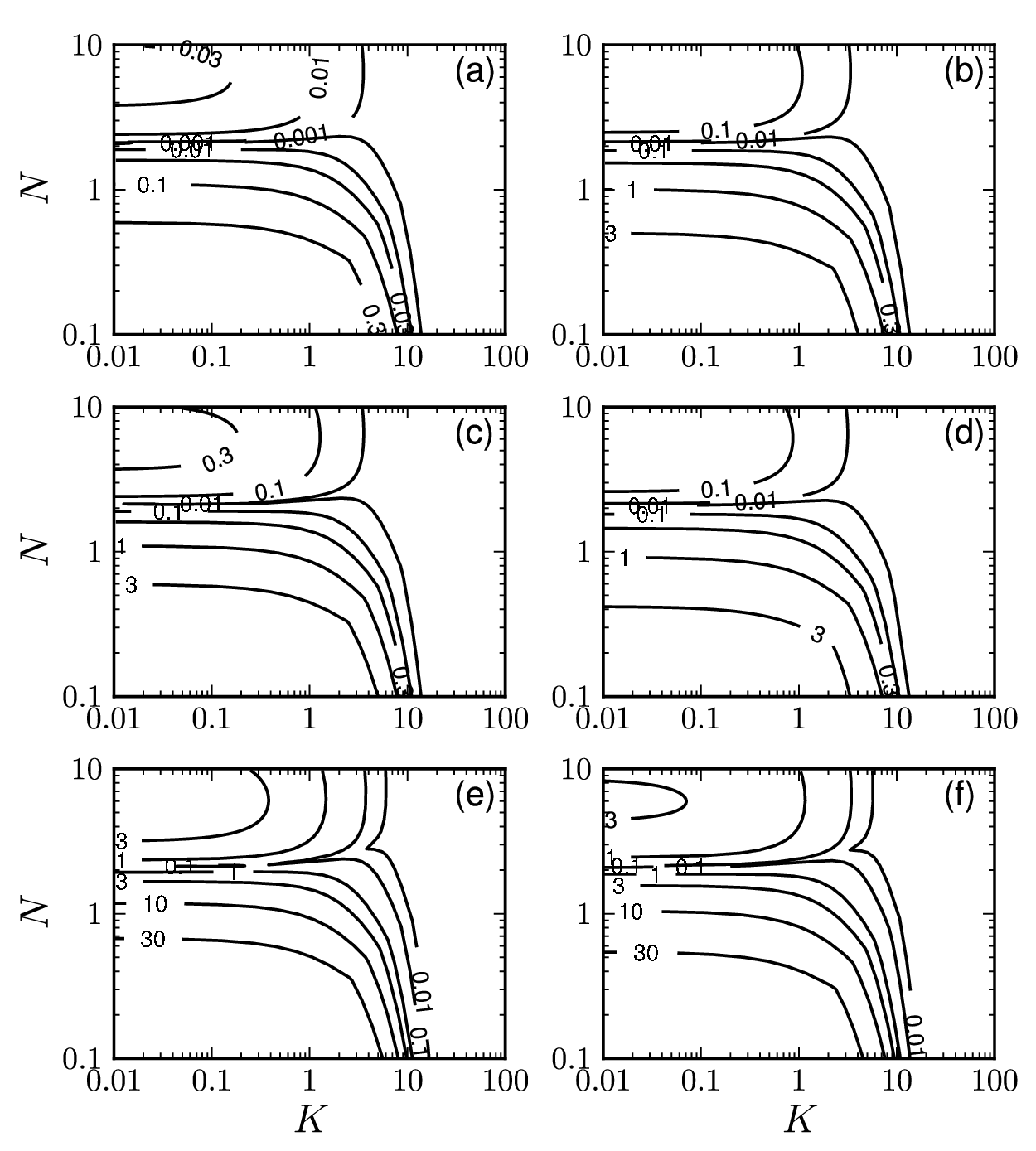}
	\caption{Contours of angle-averaged signal-to-noise ratio $\langle d \rangle$ versus normalised compressibility $K$ and Brunt-V\"ais\"al\"a frequency $N$, for four existing and planned interferometers: (a) Initial LIGO, (b) Zero-detuning, high-power Advanced LIGO, (c) Neutron star optimised Advanced LIGO, (d) Black hole optimised Advanced LIGO, (e) Conventional ET, (f) Xylophone ET.  Source parameters: $f_* = 100$ Hz, $\delta\Omega / \Omega = 2 \times 10^{-4}$, $E = 10^{-17}$, $D=1$ kpc.  The detector spectral noise densities used are $(S_h(f_*),S_h(2f_*))$:  Initial LIGO $(1.75 \times 10^{-45} \textrm{ Hz}^{-1}, 8.53 \times 10^{-46} \textrm{ Hz}^{-1})$, zero-detuning, high-power Advanced LIGO $(1.59 \times 10^{-47} \textrm{ Hz}^{-1}, 1.39 \times 10^{-47} \textrm{ Hz}^{-1})$, neutron-star-optimised Advanced LIGO $(1.18 \times 10^{-47} \textrm{ Hz}^{-1}, 9.03 \times 10^{-48} \textrm{ Hz}^{-1})$, black-hole-optimised Advanced LIGO $(3.77 \times 10^{-47} \textrm{ Hz}^{-1}, 1.84 \times 10^{-47} \textrm{ Hz}^{-1})$, conventional ET $(6.68 \times 10^{-50} \textrm{ Hz}^{-1}, 6.68 \times 10^{-50} \textrm{ Hz}^{-1})$, and xylophone ET $(1.56 \times 10^{-49} \textrm{ Hz}^{-1}, 1.12 \times 10^{-49} \textrm{ Hz}^{-1})$.} \label{fig:detectability}
\end{figure*}

Figure \ref{fig:detectability} displays contours of the average signal-to-noise ratio for Initial LIGO, Advanced LIGO (zero detuning and high laser power, neutron star optimised, and black hole optimised), and ET (conventional and xylophone) as a function of compressibility $K$ and Brunt-V\"ais\"al\"a frequency $N$.  The figure is produced for an object with $f_* = 100$ Hz, $E = 10^{-17}$, at a distance $D=1$ kpc from Earth, with radius $R=10$ km, mass $M = 1.4 \mathrm{M}_\odot$, $\rho_0 = 3 M/4\pi R^3$, and $g = GM/R^2$ (Ekman pumping occurs in a thin surface layer, where $g$ is uniform).  The step increase in angular velocity is taken to be $\delta\Omega/\Omega=2 \times 10^{-4}$, corresponding to the largest glitch observed to date \citep{mel08}.  The spectral noise densities $[S_h(f_*),S_h(2f_*)]$ used for the six detector configurations are:  Initial LIGO $(1.74 \times 10^{-45} \textrm{ Hz}^{-1}, 8.54 \times 10^{-46} \textrm{ Hz}^{-1})$, zero-detuning, high-power Advanced LIGO $(1.59 \times 10^{-47} \textrm{ Hz}^{-1}, 1.39 \times 10^{-47} \textrm{ Hz}^{-1})$, neutron-star-optimised Advanced LIGO $(1.18 \times 10^{-47} \textrm{ Hz}^{-1}, 9.03 \times 10^{-48} \textrm{ Hz}^{-1})$, black-hole-optimised Advanced LIGO $(3.77 \times 10^{-47} \textrm{ Hz}^{-1}, 1.84 \times 10^{-47} \textrm{ Hz}^{-1})$, conventional ET $(6.68 \times 10^{-50} \textrm{ Hz}^{-1}, 6.68 \times 10^{-50} \textrm{ Hz}^{-1})$, and xylophone ET $(1.56 \times 10^{-49} \textrm{ Hz}^{-1}, 1.12 \times 10^{-49} \textrm{ Hz}^{-1})$.

It is clear from Figure \ref{fig:detectability} that detectability drops off sharply for  $K > 10$.  Buoyancy prevents Ekman pumping from spinning up the whole of the stellar interior \citep{van08}.  For large stratification ($K_s \gg F N^2$), $K_s \approx K$, only a small volume of the interior is spun up and the current quadrupole is greatly reduced, with $A_{1,2} \propto e^{-2K}$.  There is little difference between the three Advanced LIGO configurations in panels (b), (c) and (d), or the two ET configurations displayed in panels (e) and (f) in Figure \ref{fig:detectability}.  All have similar sensitivity at 100 Hz.  For ET, we find $\langle d \rangle \ga 3$ for $K \la 10$ and $N \la 1$ and there is a reasonable possibility of detection.  We require smaller values, e.g. $N \la 0.5$ and $K \la 3$, to achieve $\langle d \rangle \ga 3$ with Advanced LIGO.

To generalise the results in Figure \ref{fig:detectability} to an arbitrary object, we note that $\langle d \rangle$ scales with Ekman number as $\langle d \rangle \propto E^{1/4}$ (square root of the number of cycles in the coherent integration).  For relaxation time-scales of 3 to 300 days \citep{per06}, and assuming $K=N=1$, $E$ ranges from $10^{-21}$ to $10^{-17}$, which corresponds to an order of magnitude of variation in $\langle d \rangle$.  The signal-to-noise ratio also scales with the spin parameters through the characteristic wave strain, viz.
\begin{equation}
	h_0 = 6 \times 10^{-26} \left( \frac{\delta\Omega / \Omega}{10^{-4}} \right) \left( \frac{f_*}{10^2 \textrm{ Hz}} \right)^3 \left( \frac{D}{\textrm{1 kpc}} \right)^{-1}~.\label{eq:characteristicwavestrain}
\end{equation}
The relative change in angular velocity $\delta\Omega/\Omega$ is not necessarily equal to the observed glitch size $\delta\nu/\nu$.  In a vortex unpinning model, the two quantities are related through $\delta\nu/\nu \sim (I_s/I_c)(\Delta r/R)(\delta\Omega/\Omega)$, where $I_s/I_c \sim 10^2$ is the ratio of superfluid to crust moment of inertia, and $\Delta r/R\sim10^{-6}$ is the normalised radial distance the unpinned vortices move \citep{alp86,mel10}.  Therefore, equating the observed glitch size to $\delta\Omega/\Omega$ yields a conservative estimate, given that $\delta\Omega/\Omega$ may in fact be up to $\sim 10^4$ times larger.

A coherent search synchronised to a radio ephemeris assumes that the radio and gravitational wave signals have the same phase.  This is not necessarily true.  For example, in the landmark coherent $\mathcal{F}$-statistic search for the Crab pulsar in LIGO S5 data, \citet{abb08} allowed for a fractional phase mismatch of up to $10^{-4}$.  In our multiple scales analysis, we assume by construction that the nonaxisymmetric modes are stationary in the frame rotating with the pre-glitch angular velocity and remain so throughout the Ekman pumping process.  In reality, the crust spins up to $\Omega + \delta\Omega$ and drags the axisymmetric part of the flow asymptotically to this increased angular velocity.  Whether the angular velocity of the $m\ne 0$ modes also increases during this process is unclear.  It depends on exactly how the superfluid vortices repin following a glitch and rearrange themselves in a sheared Ekman flow, which is unknown at present.

The number of templates required for a search can be estimated by modelling the frequency as $f(t) = f_0 + \dot{f}_0 t$.  For a coherent search, the difference in phase between the model and gravitational-wave signals over the integration time must satisfy $\Delta\varphi<\pi$.  For a two week integration, this corresponds to a maximum template spacing of $\delta f_0 = 3 \times 10^{-6}$ Hz and $\delta \dot{f}_0 = 4 \times 10^{-12}$ s$^{-2}$.  

During the glitch recovery, the frequency derivative is much larger than usual for an isolated pulsar spinning down electromagnetically.  We approximate $\dot{f}_0 \approx \Delta\nu / T_0$.  Conservatively, this yields $\dot{f}_0 \sim 10^{-7}$ s$^{-2}$ for a 100 Hz pulsar undergoing the largest glitch observed to date ($\Delta\nu/\nu \sim 10^{-4}$) with an unusually short relaxation period of one day.  This translates into a range of $\Delta \dot{f}_0 = 10^{-7}$ s$^{-2}$ to search over and hence $3 \times 10^4$ templates in $\dot{f}_0$.  To allow for some mismatch between the radio and gravitational wave phases we follow \citet{abb08} who searched over a window of $\pm 6 \times 10^{-3}$ Hz centred on the radio frequency, i.e. $\Delta f_0 = 1.2 \times 10^{-2}$ Hz.  Overall, therefore, a total of $\sim 10^8$ templates are required for a glitch search.

The parameters $N$, $K$ and $E$ change the shape of the signal in two ways:  the relaxation time is controlled predominantly by $E$, while the relative difference between the signals at $f_*$ and $2f_*$ (in amplitude and relaxation time) is controlled by $N$ and $K$.  Our signal-to-noise ratio estimates in Figure \ref{fig:detectability} are based on the incoherent sum of the detector response at $f_*$ and $2f_*$, so the relative phasing between $f_*$ and $2f_*$ does not affect the detectability and the number of templates required.  This would change in a more sophisticated search that combined the $f_*$ and $2f_*$ responses coherently.

To this point, we assume that radio observations provide the frequency, recovery time-scale, and trigger epoch for a glitch search.  We now consider the scenario where this information is not known, as in a blind search.  In the region of parameter space that we consider, $0.1 \le N \le 10$, $0.1 \le K \le 10$, and $10^{-20} \le E \le 10^{-8}$, the minimum band width of the Fourier-transformed wave strain is $\approx 6  \times 10^{-12} f_* $.  Hence, searching over the frequency range 1--600 Hz requires $\sim 10^{12}$ templates in $f_0$ multiplied by $\sim 10^4$ templates in $\dot{f}_0$ as discussed above.

In addition, the sky position, time of occurrence, and recovery time are unknown for a blind  search.  In a LIGO search for unknown periodic sources \citep{abb07b}, the sky is divided into 31500 patches.  The lack of an electromagnetic trigger means that the data must be searched in many blocks, starting, for example, one day apart (coherent integration over a shorter recovery time is unlikely to be detectable) and integrating over increasing lengths of time, up to the computational limit, to account for the fact that $T_0$ is unknown.  A proper estimate of the associated computational expense lies outside the scope of this paper.

\section{Constitutive properties of bulk nuclear matter}
\label{sec:nuclearproperties}

Figure \ref{fig:detectability} clearly demonstrates that the strength of the gravitational wave signal depends sensitively on the constitutive properties of bulk nuclear matter (e.g., the equation of state) and its dissipative or transport coefficients (e.g., viscosity).  We show that these properties can be inferred in principle from the detailed shape of the gravitational wave signal.  The results of this approach can be linked to terrestrial experiments, e.g. with heavy ion colliders, although there is an important distinction between $\sim$ GeV collisions of $\sim 10^2$ nucleons in a terrestrial particle accelerator and $\sim 10^{57}$ static nucleons at $\sim$ MeV energies in a neutron star.

In a real search, one seeks to extract parameters like $K$ and $N$ by fitting a template to the interferometer data in the time domain \citep{cla07,hay08}.  However, to illustrate the scientific potential of the fitting exercise, we Fourier transform $h_+(t)$ and $h_\times(t)$ and focus on the gross features of the spectrum.  We neglect the permanent fossil quadrupole (see Section~\ref{subsec:gravitationalwavestrain}) and assume that there is no interference between peaks.  The four peak amplitudes and four peak widths (at $f_*$ and $2f_*$) of $h_+(f)$ and $h_\times(f)$ provide enough information to solve for $K$, $N$, $E$, $i$, and $h_0$ by matching to the theoretical predictions in (\ref{eq:h1plus})--(\ref{eq:h2cross}).  We take ratios to eliminate $h_0$ (which depends on the unknowns $\rho_0$, $R$, $\delta\Omega$, and $D$) and focus on the remaining parameters.

\begin{figure*}
	\includegraphics{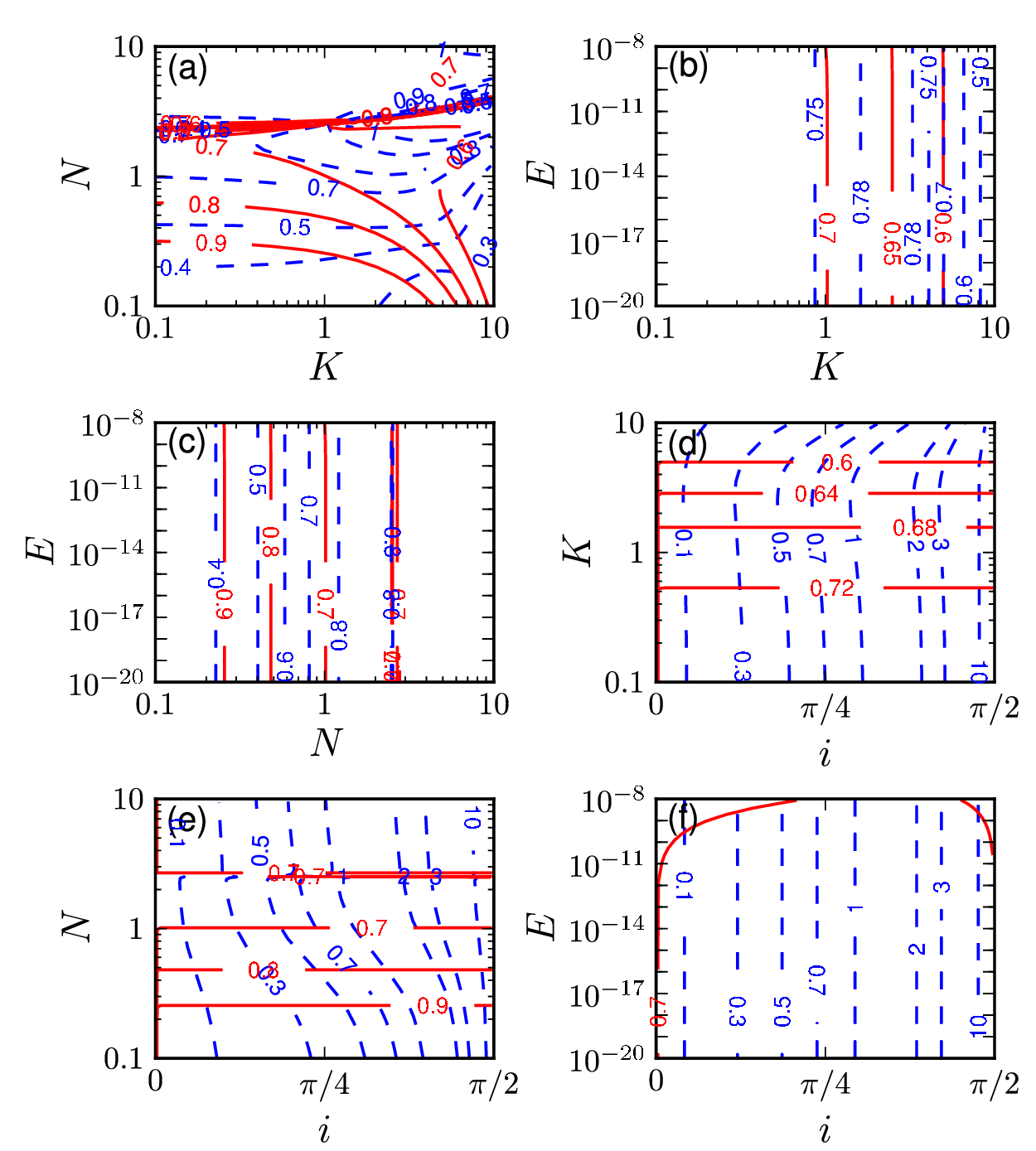}
	\caption{Ratio of fundamental and first-harmonic Fourier amplitudes $|h_+(f_*)|/|h_+(2f_*)|$ (dashed blue contours) and full widths at half maximum $\Gamma_+(f_*)/\Gamma_+(2f_*)$ (solid red contours) for six slices of the four-dimensional parameter space $(N,K,E,i)$.  In each panel, two variables are fixed, with $K = 1$, $N=1$, $E = 10^{-17}$, and $i=\pi/4$ as appropriate.}\label{fig:nuclear}
\end{figure*}

Figure \ref{fig:nuclear} displays six slices through the four-dimensional parameter space.  Contours are shown for the amplitude ratio $|h_+(f_*)|/|h_+(2f_*)|$ and the width ratio $\Gamma_+(f_*)/\Gamma_+(2f_*)$, where $\Gamma_{+,\times}(f_*)$ is the full width at half maximum of the peak in $|h_{+,\times}(f)|$ centred at $f_*$.  The figure is drawn for the parameter ranges $0.1 \le K \le 10$, $0.1 \le N \le 10$, $10^{-20} \le E \le 10^{-8}$, and $0 \le i \le \pi$.  We evaluate the first 20 terms in the infinite sums in (\ref{eq:h1plus})--(\ref{eq:h2cross}).  In those panels where $K$, $N$, $E$, and $i$ are held fixed, we use the fiducial values $K = 1$, $N=1$, $E = 10^{-17}$, and $i=\pi/4$ respectively.

The inclination angle determines the relative strength of the $m=1$ and $m=2$ modes of $h_+$ and $h_\times$ through the tensor spherical harmonic $T^{B2,lm}_{jk}$ in (\ref{eq:gwstraindefinition}).  The contours of $|h_+(f_*)|/|h_+(2f_*)|$ are nearly vertical in panels (d), (e) and (f) of Figure \ref{fig:nuclear}.  In fact, if we consider additional amplitude ratios, we can infer $i$ independently from the other parameters.  Dividing the Fourier transforms of (\ref{eq:h1plus}) by (\ref{eq:h1cross}), and (\ref{eq:h2plus}) by (\ref{eq:h2cross}), we obtain $|h_+(f_*)|/|h_\times(f_*)| = \sec i$ and $|h_+(2f_*)|/|h_\times(2f_*)| = 2 \cos i/(1+\cos^2 i)$ respectively.  These expressions overdetermine $i$, yielding its value and an independent cross-check.  The inclination angle can also be inferred from the radio or gamma-ray pulse profile and polarisation swing by assuming a particular emission model \citep{lyn88,hib01,bai09,chu10}.  The width ratios are independent of $i$; the time-scale over which the signal decays does not depend on the location of the observer.  This is illustrated in panels (d), (e) and (f) of Figure \ref{fig:nuclear}, where the contours of $\Gamma_+(f_*)/\Gamma_+(2f_*)$ are horizontal.

The compressibility $K$ and Brunt-V\"ais\"al\"a frequency $N$ are inextricably linked in the sense that they feed into both the amplitude and width ratios in a complicated manner.  However, once we determine $i$ according to the formula above, we can immediately extract $K$ and $N$ from panel (a) of Figure \ref{fig:nuclear} as the value of $E$ does not influence any of the amplitude or width ratios (see below).  By plotting contours of the measured ratios of $|h_+(f_*)|/|h_+(2f_*)|$ and $\Gamma_+(f_*)/\Gamma_+(2f_*)$ on the $K$-$N$ plane, we can read off the values of $N$ and $K$ from the intersection point of the contours.  One might be tempted to use other amplitude and width ratios as a cross-check on $K$ and $N$, or to break the degeneracy in the case of multiple intersection points.  However, most of the ratios are related trivially through the inclination angle and supply no additional information, e.g. $\Gamma_\times(f_*)=\Gamma_+(f_*), \Gamma_\times(2f_*)=\Gamma_+(2f_*)$, $|h_\times(f_+)|=\cos{i}|h_+(f_+)|$, and $|h_\times(2f_+)|=(\cos{i}+\sec{i})|h_+(2f_+)|/2$

The Ekman number $E$ is important in determining the recovery time-scale and hence the Fourier width.  It also appears in $h_0$ through $T_0$.  However, it influences all peaks in the same way and drops out of all amplitude and width ratios.  In panels (b), (c) and (f) of Figure \ref{fig:nuclear}, the amplitude and width ratio contours are vertical.  As mentioned in Section~\ref{sec:detectability}, an approximate value of $E$ can be inferred from the e-folding time of $h(t)$, as $K$ and $N$ only weakly influence this quantity.  However, if $K$ and $N$ are known, e.g. by following the procedure described in the above paragraph, we can determine $E$ from the absolute peak widths.  Finally, $h_0$ can be determined from the absolute peak amplitudes once the values of all the other parameters are known.

Future gravitational-wave measurements of the compressibility, viscosity and Brunt-V\"ais\"al\"a frequency of bulk nuclear matter can be compared to a range of terrestrial experiments and theoretical calculations.  The compressibility is commonly expressed in terms of the compression modulus $\kappa$, which is related to our normalised compressibility through $K=A m_p g R/\kappa$, where $A$ is the mean atomic number and $m_p$ the proton mass \citep{van08}.  Heavy-ion collisions and nuclear resonance experiments measure $\kappa$ \citep{stu01,vre03,pie04,har06}.  Compressibility can also be inferred from the symmetry energy measured in heavy-ion collisions or obtained through neutron-skin thickness measurements \citep{che05,li08,xu09}.  The shear viscosity is often expressed in terms of the ratio $\eta/s$, where $s$ is the specific entropy.  It is related to the Ekman number by $E=(A' k_B/m_p R^2 \Omega)(\eta/s)$ where $1 \le A' \le 2$ is the entropy per nucleon in units of Boltzmann's constant \citep{van08}.  The shear viscosity has also been measured in heavy-ion collisions \citep{adl03,ada07}.  Neutron stars are stably stratified because the concentration of charged particles increases with density but chemical equilibrium is maintained \citep{rei92}.  Stratification provides a buoyancy force proportional to the Brunt-V\"ais\"al\"a frequency squared, which has been calculated theoretically \citep{rei92,lai94,pas09}.

In Table \ref{tbl:nuclearvalues} we quote a selection of experimental and theoretical values for $K$, $N$, and $E$ under neutron star conditions.  Dimensionless values of $N$ and $E$ assume $\Omega/2\pi=100$ Hz.  In line 1, the compression modulus is inferred from the ratio of the $K^+$ multiplicity in Au+Au and C+C collisions at $\sim$ GeV energies \citep{stu01,har06}.  In line 2, the compression modulus is obtained by fitting a relativistic mean-field model to the distribution of isoscalar monopole and isovector dipole strengths of Zr and Pb \citep{vre03,pie04}.  In line 3, the compression modulus is obtained from the measured nuclear symmetry energy from isospin diffusion in heavy-ion collisions \citep{che05,li08}.  Line 4 lists the ratio of shear viscosity to specific entropy measured in Au+Au collisions at an energy of 200 GeV \citep{adl03,ada07}.  Theoretical calculations of shear viscosity by \citet{cut87} for neutron-neutron and electron-electron scattering, corresponding to the normal and superfluid states respectively, are listed in lines 5 and 6.  More exotic states, which may exist in the neutron star core, will have a different viscosity.  Line 7 lists the shear viscosity due to quark-quark scattering \citep{jai08}.  In lines 8 and 9, we quote calculated values for the Brunt-V\"ais\"al\"a frequency, the latter including centrifugal forces in a rapidly rotating star \citep{rei92,lai94,pas09}.

\begin{table*}
\setlength{\tabcolsep}{10pt}
\centering
\caption{Experimental and theoretical results for compressibility, viscosity and Brunt-V\"ais\"al\"a frequency.}
\begin{tabular}{ccccc}
\hline
Quantity & Experiment/Theory (E/T) & Result  & Dimensionless & Reference\\ \hline
\hline
$K$ & Au+Au and C+C collisions ($\sim$ GeV) (E) & $\kappa \approx 200$ MeV & $K=0.97$ & $1,2$ \\
 & nuclear resonances (E) & $\kappa \approx$ 240--270 MeV & $K=0.72$--$0.81$ &  $3,4$ \\ 
 & nuclear symmetry energy (E) & $\kappa = 210$ MeV & $K=0.93$ & $5,6$ \\ \hline
$E$ & Au+Au collisions (200 GeV) (E)& $\eta/s \approx \hbar/4\pi k_B$ & $E=8\times10^{-20}$ & $7,8$ \\
 & neutron-neutron scattering (T) & $\eta=2\times 10^{20}$ g cm$^{-1}$ s$^{-1}$ &  $E=5\times10^{-9}$ & $9$  \\
 & electron-electron scattering (T) & $\eta=6\times 10^{20}$ g cm$^{-1}$ s$^{-1}$ &  $E=1\times10^{-8}$ & $9$ \\ 
 & quark-quark scattering (T) & $\eta=5\times 10^{15}$ g cm$^{-1}$ s$^{-1}$ &  $E=1\times10^{-13}$ & $10$  \\ \hline
$N$ & chemical composition (T) & $N_* \sim$ 500 s$^{-1}$ & $N= 0.8$ & $11,12$ \\
 & centrifugal correction (T) & $N = 0.32$-$0.84$ & $N = 0.32$-$0.84$ & $13$ \\
\hline
\multicolumn{5}{l}{(1)~\citet{stu01}, (2)~\citet{har06}, (3)~\citet{vre03}, (4)~\citet{pie04}, (5)~\citet{che05}, } \\
\multicolumn{5}{l}{(6)~\citet{li08}, (7)~\citet{adl03}, (8)~\citet{ada07}, (9)~\citet{cut87}, (10)~\citet{jai08},} \\
\multicolumn{5}{l}{(11)~\citet{rei92}, (12)~\citet{lai94}, (13)~\citet{pas09}} \\
\end{tabular}
\label{tbl:nuclearvalues}
\end{table*}

\section{Conclusions}
\label{sec:conclusion}

In this paper, we calculate analytically the gravitational radiation emitted during the post-glitch recovery phase by the nonaxisymmetric Ekman flow excited by a glitch.  The calculation is done in the context of an idealised, cylindrical star with a uniform viscosity, compressibility, and stratification length-scale.  We compute the signal-to-noise ratio for current- and next-generation long-baseline interferometers and find the following promising result:  for a large glitch ($\delta\Omega/\Omega =10^{-4}$) from a neutron star $D=1$ kpc from Earth and spinning at $f_* = 100$ Hz, the angle-averaged signal-to-noise ratio $\langle d \rangle$ exceeds three for $N\la 0.5$, $K\la 10$, and $E\sim10^{-17}$ with Advanced LIGO and $N\la 1$, $K\la 10$, and $E\sim10^{-17}$ with ET.

Perhaps the most obvious shortcoming of our idealised model is its cylindrical geometry.  There is a noble history of using a cylinder to model spherical astronomical objects and also in classical geophysical studies of the Earth \citep{ped67, wal69, abn96, van08}, because it admits analytic solutions, which in general have not yet been found for a sphere.  We ignore magnetic fields for simplicity, although they are large in neutron stars \citep{cut02}, interact with the superfluid \citep{men98}, and therefore modify Ekman pumping.  We model the interior of a neutron star as a single Navier-Stokes fluid, whereas in reality it is a multi-component superfluid, consisting of superfluid neutrons and superconducting protons which interact with each other via mutual friction and entrainment \citep[e.g.][]{lat04,and06}.  The spin-up process in a coupled multi-component fluid of this kind, in the presence of gravitational stratification and compressibility, is an unsolved and difficult problem.

In our model, the crust accelerates instantaneously from $\Omega$ to $\Omega + \delta\Omega$ and remains at this higher angular velocity.  A more realistic model would conserve total angular momentum by solving self-consistently for the response of the crust to the viscous back-reaction torque \citep{van10}.  In the context of the present model, we can approximate this effect crudely by replacing the glitch size $\delta\Omega/\Omega$ at $t=0$ with the permanent frequency jump after the recovery ceases.  None of the conclusions change qualitatively.

Understanding the glitch mechanism remains an unsolved problem.  Glitch waiting times are exponentially distributed and their sizes fit a power law \citep{mel08}, indicative of inhomogeneous collective behaviour on large scales, e.g. vortex avalanches.  In contrast, nuclear structure calculations suggest that the area density of pinning sites (e.g. lattice defects) is much greater than the area density of vortices \citep{jon02a,don03}, suggesting that the system is homogeneous on large scales (pinned Abrikosov array).  The gravitational wave signal calculated here helps to discriminate between these two views, as it is a measure of the internal nonaxisymmetry.  From a simple, random walk argument, the largest relative glitch size that arises from vortex movement in a star containing $n$ vortices is $\delta\Omega/\Omega\sim n^{-1/2}$.  If the value of $\delta\Omega/\Omega$ inferred from a gravitational-wave detection approaches this maximum, it is safe to infer that large-scale inhomogeneities are present.  Note that we take $C_m=1$ in Section~\ref{subsec:chooseBCICforglitch}.  However, if only a fraction of the internal flow is nonaxisymmetric, $C_m$ should be reduced in proportion.  In vortex unpinning models $\delta\Omega/\Omega$ can be up to four orders of magnitude larger than the observed glitch size (see Section~\ref{subsec:detectabilityestimate}), leaving considerable scope to get detectable gravitational-wave signals.

Vortex unpinning theories of glitches rely on the build up of a lag between the crust, which spins down electromagnetically, and the superfluid, whose rotation is fixed by the number of vortices, until a glitch is triggered.  We know that the lag does not disappear completely after the glitch (i.e. co-rotation is not restored) because a reservoir effect (i.e. glitch size $\propto$ waiting time) is not observed in glitch data \citep{won01}; only a small, random fraction of the lag relaxes during a single event, and that fraction is determined by the microscopic history of the system, as in any avalanche process.  In this model, we assume conservatively that the crust and fluid co-rotate before the glitch.  However, in the more realistic scenario just described, there is ongoing differential rotation between crust and core, suggesting that glitching pulsars may continuously emit gravitational radiation.

Another possibility leading to a continuous gravitational wave signal beyond just the post-glitch recovery period is the `fossil flow' discussed in Section~\ref{sec:gravitationalwavecalculation}.  Stratification prevents Ekman pumping from spinning up the whole interior, leaving a remnant of the initial nonaxisymmetric flow untouched.  This flow emits gravitational radiation until damped over the much longer diffusion time-scale.  If so, we may be able to extend the coherent integration time beyond the recovery time-scale, increasing the likelihood of detection.  Even more intriguing is the possibility that any neutron star which has experienced differential rotation in its past retains some part of this fossil flow for $\ga 10^3$ years, thereby bearing an imprint of the star's formation and rotation history.  We plan to study the matter fully in a following paper.

For a typical neutron star at a distance of 1 kpc, the signal-to-noise calculations in Section~\ref{sec:detectability} argue that there is a reasonable chance interferometers like Advanced LIGO or ET will detect the largest glitches.  The outlook is more optimistic if we consider nearby `dark' neutron stars.  For the estimated galactic population of $\sim 10^9$ neutron stars (cf., $\sim 1800$ radio pulsars discovered to date),  recent Monte-Carlo simulations predict the closest objects are located $\sim 8$ pc from Earth \citep{ofe09}.  At this distance, Initial LIGO is able to detect the largest glitches with $\langle d \rangle \ga 3$ for $N \la 1$ and $K \la 10$ and Advanced LIGO is sensitive to smaller glitches with $\delta\Omega/\Omega \ga 10^{-6}$.  However, the signal frequency, glitch epoch, and sky position are unknown electromagnetically, so searching for `dark' glitches is a difficult proposition.  None the less, our results suggest cautious optimism about the chances of detecting a glitching (or otherwise differentially rotating) neutron star with the next generation of gravitational-wave interferometers.

\section*{Acknowledgements}
We thank the anonymous referee for their helpful comments and suggestions.  MFB and CAVE acknowledge the support of Australian Postgraduate Awards.

\bibliographystyle{mn2e}
\bibliography{glitchGW}

\appendix
\section{Simplifying $\vx\bdot\textrm{curl}(\rho\vv)$}\label{sec:appendixcurlrhov}  
It is straightforward to evaluate $S^{lm}$ by substituting (\ref{eq:vrsolution})--(\ref{eq:psolution}) directly into (\ref{eq:currentquadrupoledefinition}).  However, the calculation is easier and more transparent if we first simplify the integrand in (\ref{eq:currentquadrupoledefinition}) to depend only on $\delta p$.  Expanding according to $\rho \mapsto \rho^0 + \delta\rho$, $\vv \mapsto \vv^0 + \bmath{\delta v} = r\Omega \bmath{e_{\phi}} + \bmath{\delta v}$, we express the integrand to first order as
\begin{equation}
	\vx \bdot \nabla \btimes (\rho^0 \vv^0 + \delta\rho \vv^0 + \rho^0 \bmath{\delta v})~.
	\label{eq:curlrhov1}
\end{equation}
The first term in (\ref{eq:curlrhov1}) is independent of time.  It does not emit gravitational radiation, so we discard it.  The second term in (\ref{eq:curlrhov1}) reads
\begin{eqnarray}
	\vx \bdot \nabla \btimes (\delta\rho \vv^0) &=& \Omega \left( 2 z + r z \frac{\partial}{\partial r} - r^2 \frac{\partial}{\partial z} \right) \delta\rho~,\label{eq:appendix1}\\
	&=& \frac{\Omega}{g} \left( r^2 \frac{\partial^2}{\partial z^2} - rz \frac{\partial^2}{\partial r \partial z} - 2 z \frac{\partial}{\partial z} \right) \delta p~.\label{eq:RHSterm1}
\end{eqnarray}
To move from (\ref{eq:appendix1}) to (\ref{eq:RHSterm1}) we use the Navier-Stokes equation (\ref{eq:navier-stokes}) to first order in Rossby number and zeroth order in Ekman number.  In the rotating frame and neglecting the centrifugal term, as in Section~\ref{sec:ekmanflow}, it reads
\begin{equation}
	2 \rho^0 (\vOmega\btimes\bmath{\delta v}) = - \nabla \delta p - \vg \delta\rho~,
	\label{eq:navier-stokesleadingorder}
\end{equation}
from which we obtain
\begin{equation}
	\delta\rho = -\frac{1}{g} \frac{\partial \delta p}{\partial z}~. \label{eq:appendix2}
\end{equation}
The third term in (\ref{eq:curlrhov1}) can be rewritten in a similar way.  From (\ref{eq:navier-stokesleadingorder}), we find
\begin{equation}
	2\Omega \rho^0 \bmath{e_{z}}\btimes (\bmath{e_{z}} \btimes \bmath{\delta v}) = \nabla \btimes (\delta p \bmath{e_{z}})~,
\end{equation}
where $(\bmath{e_{r}},\bmath{e_{\phi}},\bmath{e_{z}})$ are the basis vectors in cylindrical coordinates.  Noting that $\delta v_z^0=0$, as the axial flow is $O(E^{1/2})$, we are left with
\begin{equation}
	\rho^0 \bmath{\delta v} =- \frac{1}{2\Omega} \nabla\btimes(\delta p \bmath{e_{z}})~,
	\label{eq:curlrhov2}
\end{equation}
and the third term in (\ref{eq:curlrhov1}) is
\begin{equation}
	\vx\bdot\nabla\btimes(\rho^0 \bmath{\delta v}) = \frac{1}{2\Omega}\left( z\nabla^2 - r\frac{\partial^2}{\partial r \partial z} - z\frac{\partial^2}{\partial z^2} \right) \delta p ~.\label{eq:RHSterm2}
\end{equation}
Combining (\ref{eq:RHSterm1}) and (\ref{eq:RHSterm2}), and replacing $\partial^2/\partial\phi^2$ by $-m^2$, we arrive at
\begin{equation}
	\vx\bdot\nabla\btimes(\rho \vv) = \Bigg[ \frac{1}{2\Omega} \left(z \frac{\partial^2}{\partial r^2} + \frac{z}{r}\frac{\partial}{\partial r} - \frac{z m^2}{r^2} - r\frac{\partial^2}{\partial r \partial z}\right) + \frac{\Omega}{g} \left( r^2 \frac{\partial^2}{\partial z^2} - rz\frac{\partial^2}{\partial r \partial z} - 2 z \frac{\partial}{\partial z} \right) \Bigg] \delta p^0~.
	\label{eq:curlresult}
\end{equation}

There is a subtle issue around neglecting the centrifugal correction to (\ref{eq:navier-stokesleadingorder}), which is of order $F$.  If we evaluate $S^{lm}$ by substituting (\ref{eq:vrsolution})--(\ref{eq:psolution}) directly into (\ref{eq:currentquadrupoledefinition}), we implicitly include centrifugal terms in $\vx\bdot\nabla\btimes(\rho \vv)$ (by virtue of failing to exclude them explicitly).  This approach is internally inconsistent, because centrifugal terms of this order are excluded from the flow fields (\ref{eq:vrsolution})--(\ref{eq:psolution}) following the assumption in Section~\ref{subsec:spinupflow} leading to (\ref{eq:rhoequilsolution}) and (\ref{eq:pequilsolution}).  It is therefore preferable to evaluate (\ref{eq:currentquadrupoledefinition}) for $S^{lm}$ using (\ref{eq:curlresult}), so that the centrifugal correction to the zeroth-order structure is consistently excluded from both the flow fields and $S^{lm}$.

\section{Beam pattern functions}\label{sec:appendixbeampatternfunctions}

The complete expressions for the beam pattern functions are \citep{jar98},
\begin{eqnarray}
	F_+(t) &=& \sin\zeta [ a(t) \cos 2\psi + b(t) \sin 2\psi ]~,\label{eq:appendixFplus}\\
	F_\times(t) &=& \sin\zeta [ b(t) \cos 2\psi - a(t) \sin 2\psi ]~,\label{eq:appendixFcross}
\end{eqnarray}
with
\begin{eqnarray}
	a(t) &=& \frac{1}{16} \sin 2\gamma (3-\cos 2\lambda)(3-\cos 2\delta) \cos[2(\alpha-\phi_r-\Omega_r t)] - \frac{1}{4} \cos 2\gamma \sin\lambda (3-\cos 2\delta) \sin[2(\alpha-\phi_r-\Omega_r t)] \nonumber\\
	&&\;+\frac{1}{4} \sin 2\gamma \sin 2\lambda \sin 2\delta \cos[\alpha-\phi_r-\Omega_r t] - \frac{1}{2} \cos 2\gamma \cos\lambda \sin 2\delta \sin[\alpha-\phi_r-\Omega_r t] + \frac{3}{4} \sin 2\gamma \cos^2\lambda \cos^2\delta~,\label{eq:appendixat}\\
	b(t) &=& \cos 2\gamma \sin\lambda \sin \delta \cos[2(\alpha-\phi_r-\Omega_r t)] + \frac{1}{4} \sin 2\gamma (3- \cos 2\lambda) \sin\delta \sin[2(\alpha-\phi_r-\Omega_r t)]\nonumber\\
	&&\;+ \cos 2\gamma \cos\lambda \cos\delta \cos[\alpha-\phi_r-\Omega_r t] + \frac{1}{2} \sin 2\gamma \sin 2\lambda \cos\delta \sin[\alpha-\phi_r-\Omega_r t]~.\label{eq:appendixbt}
\end{eqnarray}
The right ascension and declination of the gravitational wave source are given by $\alpha$ and $\delta$ respectively, and $\psi$ is the polarisation angle.  The latitude of the detector is denoted by $\lambda$, $\Omega_r$ is the angular velocity of the Earth, and $\phi_r$ is the diurnal phase of the Earth.  The angle counterclockwise between East and the bisector of the interferometer arms is $\gamma$, and the angle between the arms of the interferometer is $\zeta$.  We average over $\alpha$, $\delta$ and $\psi$ according to \citep{jar98}
\begin{equation}
	\langle ... \rangle_{\alpha, \delta, \psi} = \frac{1}{2\pi} \int_0^{2\pi} \dd \alpha \times \frac{1}{2} \int_{-1}^1 \dd (\sin\delta) \times \frac{1}{2\pi} \int_0^{2\pi} \dd \psi \; (...) ~.
\end{equation}

We evaluate $\langle \int_0^{T_0}\dd t F_+^2\rangle_{\alpha,\delta,\psi}$ and $\langle \int_0^{T_0}\dd t F_\times^2\rangle_{\alpha,\delta,\psi}$ for use in Section~\ref{sec:detectability}.  Averaging (\ref{eq:appendixFplus}) and (\ref{eq:appendixFcross}) over $\psi$, we obtain
\begin{equation}
	\left\langle \int_0^{T_0} \dd t F_+^2 \right\rangle_{\psi} = \left\langle \int_0^{T_0} \dd t F_\times^2 \right\rangle_{\psi} =  \frac{1}{2} \sin^2\zeta \int_0^{T_0} \dd t \left([a(t)]^2 + [b(t)]^2\right)~.\label{eq:appendixFcross1}
\end{equation}
All the dependence on $\alpha$ and $\delta$ is contained in $a(t)$ and $b(t)$.  After some straightforward but lengthy algebra, we find that the dependence on all other angles drops out, leaving
\begin{equation}
	\left\langle [a(t)]^2 + [b(t)]^2 \right\rangle_{\alpha,\delta} = \frac{2}{5}~.\label{eq:appendixabresult}
\end{equation}
Substituting (\ref{eq:appendixabresult}) into (\ref{eq:appendixFcross1}) and evaluating the now trivial time integration we obtain the result stated in equation (\ref{eq:averagebeampatternfunction}),
\begin{equation}
	\left\langle \int_0^{T_0} \dd t F_+^2 \right\rangle_{\alpha, \delta, \psi} = \left\langle \int_0^{T_0} \dd t F_\times^2 \right\rangle_{\alpha, \delta, \psi} = \frac{T_0}{5} \sin^2\zeta~.
\end{equation}

\bsp
\label{lastpage}

\end{document}